\documentclass[prd,aps,a4paper,eqsecnum,floatfix,twocolumn,nofootinbib]{revtex4}  

\usepackage{graphicx} 
\usepackage{amsmath,amsfonts,amssymb}
\usepackage{placeins}
\usepackage{appendix}
\usepackage[caption=false,justification=justified]{subfig}

\newcommand{\beq}{\begin{equation}}
\newcommand{\eeq}{\end{equation}}
\newcommand{\bea}{\begin{eqnarray}}
\newcommand{\eea}{\end{eqnarray}}
\newcommand{\be}{\begin{equation}}      
\newcommand{\ee}{\end{equation}}
\def\nn{\nonumber}

\begin{document}

\title{Analytic self-force effects on radial infalling particles in the Schwarzschild spacetime: the radiated energy}

\author{Donato Bini$^{1}$, Giorgio Di Russo$^{2}$}
  \affiliation{
$^2$Istituto per le Applicazioni del Calcolo ``M. Picone,'' CNR, I-00185 Rome, Italy\\
$^3$School of Fundamental Physics and Mathematical Sciences, Hangzhou Institute for Advanced Study, UCAS, Hangzhou 310024, China\\
}

\date{\today}

\begin{abstract}
We compute, at the first self force accuracy level, the radiated energy from a radially infalling particle released from rest in a Schwarzschild spacetime.
We examine both the cases of a scalar particle and that of a massive particle, in the context of gravitational perturbations.
Our findings are accompanied by Post-Newtonian checks. In spite of the specific interest for this kind of computations, we outline the building blocks for future higher-order Post-Newtonian computations as well as for extending these results to other interesting situations out of the black hole case.
\end{abstract}

\maketitle
\section{Introduction}
\label{Intro}

Black hole perturbations due to fields of any spin-weight, e.g., scalar ($s=0$), electromagnetic ($s=1$), gravitational ($s=2$), Rarita-Schwinger ($s=3/2$), etc. have been largely studied, especially in the case of black hole spacetimes (Schwarzschild and Kerr), either numerically and analytically, see e.g. the reviews \cite{Sasaki:2003xr,Poisson:2003nc,Barack:2018yvs}. 
They crucially depend on the orbit of the particle (with proper type charge and mass) associated with the perturbation which sources the corresponding wave equation, scalar, electromagnetic (em) and gravitational (grav)
\[
\begin{array}{ll}
{\rm scalar} &  \rho=q\int \frac{d\tau}{\sqrt{-g}} \delta^{(4)}(x-x_p(\tau))\,, \cr
{\rm em} &   J^\alpha= q_{\rm em}\int \frac{d\tau}{\sqrt{-g}} u^\alpha \delta^{(4)}(x-x_p(\tau))\,, \cr
{\rm grav} &   T^{\alpha\beta}= m \int \frac{d\tau}{\sqrt{-g}} u^\alpha u^\beta\delta^{(4)}(x-x_p(\tau))\,, \cr
\end{array}
\]
where $x^\alpha=x^\alpha_p(\tau)$ denote the parametric equations of the orbit.
We currently have a satisfying analytical knowledge of effects associated with particles at rest, in a circular equatorial motion and, in general for bound equatorial motion with  small eccentricity (ellipticlike motions) as well as for unbound equatorial motion with large eccentricity (hyperboliclike motions).
Unfortunately, we do know much analytical perturbations due to radially infalling particles and the corresponding radiative losses.
Indeed, most existing works on energy loss from particles radially falling into a Schwarzschild black hole are semi-analytic or numerical, and date back to the 1970s, starting with the pioneering papers by Zerilli \cite{Zerilli:1970wzz} and by Davis, Ruffini, Press, Price and Tiomno \cite{Davis:1971gg,Davis:1972ud,Ruffini:1973ky}, followed by more recent developments \cite{Cardoso:2002jr,Mitsou:2010jv,Oliveira:2018pnx}.
From the two-body perspective, radial infall is also referred to as a "head-on collision" or a "direct plunge." Note that the multipolar expansion is compared with numerical studies in Refs. \cite{Anninos:1993zj,Simone:1995qu}.

Here, for the first time, we  compute analytically the radiated energy by scalar or massive particles within the framework of gravitational self force theory \cite{Barack:2018yvs}.
This study has not only an academic interest and cannot be labelled as an exercise based on previously developed formalisms/approaches/results, because a particle falling into a black hole will enter soon a strong field regime region, where most of the tools available in perturbation theory are no longer valid, differently from a (massive) particle in quasi-circular motion or along a hyperboliclike trajectory which can spend most of the time in a weak field region (and consequently may lead to computable, highly accurate effects in Post-Newtonian sense).
Clearly this new situation requires a more careful analysis, with eventual inclusion of cut-off integrals to avoid unphysical divergencies.
In this sense, a stimulating study is outlined in Ref. \cite{Barausse:2021xes}. 

Therefore, aiming at filling the mentioned \lq\lq analytic" gap existing in the literature, we first analyze the simplest case of scalar field and then discuss the more interesting case of gravitational perturbations in the context of gravitational self force. All these studies, when possible, will be paired with Post-Newtonian calculations which  validate the final analytic result.

As it often happens in self force, the scalar field case  contains already the main technical difficulties and it has to be considered preliminary to the others. 
We will start discussing in detail the   underlying  difficulties of an analytic computation in the region where the field is not so strong to allow for Post-Newtonian (PN) analysis. 
Viceversa, the strong field region requires $G\gg 1$ and it  is mainly accessible by numerical relativity. However, the leading order behaviour of the power spectrum was already obtained in the seminal paper of Zerilli \cite{Zerilli:1970wzz}, showing  a decreasing  exponential behavior (at large $GM\omega$) of the  energy loss at infinity. 
This topic will be explored in future (still ongoing) works aiming at obtaining systematically (at least) the subleading terms in such a regime.
A numerical update for this computation can be found in Ref. \cite{Mitsou:2010jv}, where  an estimation of the quasi normal modes (QNM) in the Regge-Wheeler-Zerilli approach is also given.
Being the radial fall trajectory approaching the black hole horizon at the end of the process, these computations should mimic the merging of the two black  holes, and consequently the \lq\lq ringdown" with emission of the characteristic QNM frequencies.
The latter topic, together with tidal deformations and wave amplification factors,  received much attention recently because of the gravitational wave signal GW250114 \cite{LIGOScientific:2025rid} recently detected with a particularly high signal-to-noise ratio, 
and the technology to determine numerically or analytically them is well established (starting from pioneering works, see e.g. Ref. \cite{Chandrasekhar:1975zza}, up to recent literature based on CFT techniques \cite{Berti:2009kk,Aminov:2020yma}), even out the black hole context \cite{Bianchi:2021xpr,Bianchi:2021mft,Bianchi:2022qph,Bianchi:2023sfs,DiRusso:2024hmd,Bena:2024hoh,Cipriani:2024ygw,Fioravanti:2021dce,DiRusso:2025qpf}.

In the present paper, after the preliminary case of scalar field perturbations, we extend the discussion to the more general situation of gravitational perturbations. 

We show explicit expressions for the energy loss in the time domain and in the Fourier domain (namely,  the main accomplishments of the present work), with some checks coming from Multipolar-Post-Minkowskian (MPM) formalism and Post-Newtonian (PN) theory.

In this work we use the mostly positive metric signature ($-+++$) and set units such that $G=1=c$, unless differently specified.

\section{Radial infalling geodesics on a Schwarzschild spacetime}

Let us consider the Schwarzschild metric in standard Schwarzschild coordinates
\bea
ds^2&=&-f(r)dt^2+\frac{dr^2}{f(r)}+r^2\left(d\theta^2+\sin^2\theta d\phi^2\right)\,,\qquad
\eea
where
\beq
f(r)=1-\frac{2M}{r}\,.
\eeq
Timelike equatorial ($\theta=\frac{\pi}{2}$) and radially infalling geodesics at $\phi=0$ (parametrized by the proper time $\tau$) have parametric equations $x^\alpha=x_p^\alpha(\tau)$ (and four velocity $U^\alpha=\frac{dx_p^\alpha}{d\tau}$) such that 
\beq
\label{dt_dtau}
\frac{dt_p}{d\tau}=\frac{ E}{f(r_p)}\,,
\eeq
with the radial motion described by the following equation
\beq
\label{dr_dt}
\frac{dr_p}{dt}=-\left(1-\frac{2M}{r_p}\right)\frac{\sqrt{E^2-1+\frac{2M}{r_p}}}{E}\,.
\eeq
Here $E$ denotes the energy per unit mass of the particle while the angular momentum $L=0$ for a radial fall.

\begin{figure}
\includegraphics[scale=0.8]{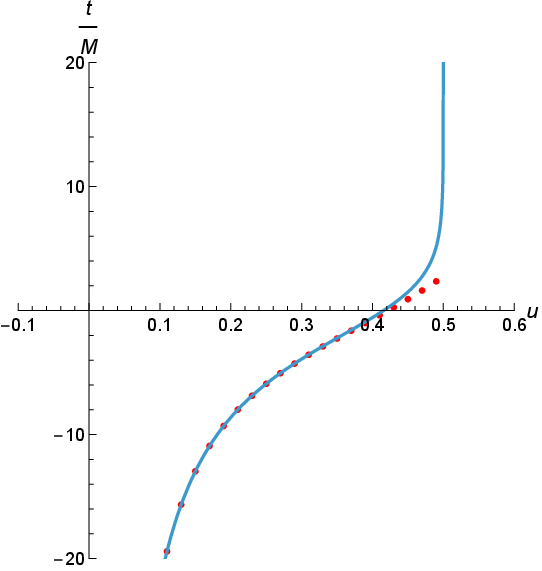}
\caption{\label{fig:1} The behavior of $t$ (in units of $M$) vs $u\in (0,\frac12)$. The falling process starts at $t=-\infty$ at radial infinity and ends at the horizon at $t=+\infty$. The superposed points (red online) correspond to the PN approximated solution which includes contributions up to $O(\eta^{16})$. The two plots start keeping away at $u\approx 0.4$ meaning the non reliability of the PN approximation at this radius (close to the horizon, $u=\frac12$).}
\end{figure}

We will limit our considerations to motion starting from rest ($E=1$) at spatial (radial) infinity, 
\beq
\label{dr_dt2}
\frac{dr_p}{dt}=-f(r_p) \sqrt{\frac{2M}{r_p}}
\,.
\eeq
Introducing the (dimensionless) inverse radial variable $r=M/u$ in Eq. \eqref{dr_dt2} the exact solution of Eq. \eqref{dr_dt} (solved by using $u$ as a parameter along the orbit instead of the proper time or the coordinate time) reads
\beq
t_p(u)=M \left[-\frac{\sqrt{2} (1+6u)}{3u^{3/2}}+2 \log\left(\frac{1+\sqrt{2u}}{1-\sqrt{2u}}\right)\right]\,.
\eeq
Fig. \ref{fig:1} shows the behavior of $t$ (in units of $M$) vs $u\in (0,\frac12)$, both in the exact case (solid curve) and in the PN approximated one (doted curve). 

When using $t$ as a parameter (as we will do systematically below), instead, the above equation can be solved in PN sense  as follows
\bea
\label{r_p_di_t}
r_p(t)&=&\alpha  (-t)^{2/3}-\frac{8}{9} \alpha ^3 \eta ^2+\frac{16 \alpha ^5 }{27
   (-t)^{2/3}}\eta ^4\nonumber\\
&+& \frac{640 \alpha ^7 }{2187 (-t)^{4/3}}\eta
   ^6-\frac{256 \alpha ^9 }{10935 t^2}\eta
   ^8\nonumber\\
&{-}&\frac{83968 \alpha ^{11} }{413343 (-t)^{8/3}}\eta
   ^{10}{-}\frac{3948544 \alpha ^{13} }{23914845 (-t)^{10/3}}\eta
   ^{12}\nonumber\\
&+&\frac{3080192 \alpha ^{15} }{122762871 t^4}\eta
   ^{14}+\frac{18389663744 \alpha ^{17}
   }{97931290275
   (-t)^{14/3}}\eta ^{16}\nonumber\\
&+&O(\eta^{17})\,,
\eea
where $\eta=\frac{1}{c}$ is a place-holder for PN expansion and
\beq
\label{alpha_def}
\alpha=\left(\frac{9M}{2}\right)^{\frac{1}{3}}\,.
\eeq
Equivalently, Eq. \eqref{r_p_di_t} can be cast in the following form
\bea
\label{r_p_di_t2}
r_p(\zeta)&=&\alpha  t^{2/3}\left[1-\frac{8}{9} \zeta^2  +\frac{16  }{27
   }\zeta ^4+\frac{640}{2187}\zeta
   ^6\right.\nonumber\\
&{-}&\frac{256 }{10935}\zeta
   ^8{-}\frac{83968}{413343 }\zeta
   ^{10}{-}\frac{3948544}{23914845}\zeta
   ^{12}\nonumber\\
&+&\frac{3080192}{122762871}\zeta
   ^{14}+\frac{18389663744
   }{97931290275
   }\zeta ^{16}\nonumber\\
&+&\left. O(\zeta^{17})\right]\,,
\eea
where
\beq
\zeta=\frac{\alpha \eta}{(-t)^{1/3}}\,.
\eeq
The parametric equations of the perturbing particle orbit  are the main ingredient needed to study non-homogeneous wave equations in the Schwarzschild spacetime.
In the next Section III below we will start considering scalar perturbations, whereas starting from Section IV we will consider the more interesting situation of gravitational perturbations.


\section{Scalar wave equation}

Let us 
consider  the scalar wave equation
\beq
\label{eq_fund}
\Box \psi=-4\pi \rho\,,
\eeq
where
\beq
\Box \psi=\frac{1}{\sqrt{-g}}\partial_\mu (\sqrt{-g}g^{\mu\nu}\partial_\nu  \psi)\,,
\eeq
with $\sqrt{-g}=r^2\sin\theta$, and the source energy density  given by
\beq
\rho=q\int \frac{d\tau}{\sqrt{-g}} \delta^{(4)}(x-x_p(\tau))\,.
\eeq

In the present case of radial falling from rest at infinity the density $\rho$  reads
\bea
\rho&=& q\int \frac{d\tau}{r^2} \delta(t-t_p(\tau))\delta(r-r_p(\tau))\delta(\theta-\frac{\pi}{2})\delta(\phi)\nonumber\\
&=& \frac{q}{r_p^2(t)U^t} \delta(r-r_p(t))\delta(\theta-\frac{\pi}{2})\delta(\phi)\nonumber\\
&=&  \frac{qf(r_p(t))}{r_p^2(t)} \delta(r-r_p(t))\sum_{lm}Y_{lm}(\theta, \phi)Y_{lm}^*(\frac{\pi}{2}, 0)\,,\qquad
\eea
where we have chosen to parametrize the orbit  by the coordinate time $t$, expressed the product $\delta(\theta-\frac{\pi}{2})\delta(\phi)$ in (scalar) spherical harmonics and performed the integral over $\tau$  by using the relation
\beq
\delta(t-t_p(\tau))=\frac{\delta(\tau-t)}{U^t(t)}=f(r_p(t)) \delta(\tau-t) \,.
\eeq
Separation of the angular variables and Fourier transforming the time variable implies
\bea
\label{rho_t2}
\rho(t,r,\theta,\phi) 
&=&  \sum_{lm}\int \frac{d\omega}{2\pi}e^{-i\omega t}\widehat T_{lm}(\omega ,r) Y_{lm}(\theta, \phi)\,, \qquad\nonumber\\
&=& \sum_{lm}T_{lm}(t,r)Y_{lm}(\theta, \phi)\,,
\eea
with
\bea
T_{lm}(t,r)&=&\frac{qf(r_p(t))}{r_p^2(t)} \delta(r-r_p(t))Y_{lm}^*(\frac{\pi}{2}, 0)\nonumber\\
&=& \int \frac{d\omega}{2\pi}e^{-i\omega t}\widehat T_{lm}(\omega ,r)\,.
\eea
Therefore,
\bea
\widehat T_{lm}(\omega ,r)&=&q\int dt e^{i \omega t } \frac{ f(r_p(t))}{r_p^2(t) } \delta(r-r_p(t)) Y_{lm}^*(\frac{\pi}{2}, 0)\nonumber\\
&=&q Y_{lm}^*(\frac{\pi}{2}, 0){\mathcal F}(r,\omega)\,,
\eea
with
\beq
{\mathcal F}(\omega,r)=\int dt e^{i \omega t }\frac{ f(r)}{r^2}  \delta(r-r_p(t))\,.
\eeq

Let us look for solutions $\psi$ of the scalar wave equation of the form
\beq
\label{psi_exp0}
\psi(t,r,\theta,\phi)=\sum_{lm}\int \frac{d\omega}{2\pi}e^{-i\omega t}R_{lm\omega}(r)Y_{lm}(\theta,\phi)\,.
\eeq
Eq. \eqref{eq_fund} then implies
\begin{widetext}
\bea
\Box \psi &=&\sum_{lm}\int\frac{d\omega}{2\pi} e^{-i \omega t}Y_{lm}(\theta,\phi) \left[f(r)R_{lm\omega}''+\frac{2(r-M)}{r^2}R_{lm\omega}'+\left(\frac{\omega^2}{f}  - \frac{L}{r^2}\right) \right]\nonumber\\
&=& -4\pi  \sum_{lm}\int \frac{d\omega}{2\pi}e^{-i\omega t}\widehat T_{lm}(\omega ,r) Y_{lm}(\theta, \phi)Y_{lm}^*(\frac{\pi}{2}, 0)\,,
\eea
(here $L=l(l+1)$) that is
\beq
\label{rad_eq_gen2}
R_{lm\omega}''+\frac{2(r-M)}{r^2f(r)}R_{lm\omega}'+\frac{1}{f}\left(\frac{\omega^2}{f}  - \frac{L}{r^2}\right)R_{lm\omega}= -4\pi\frac{1}{f} \widehat T_{lm}(\omega ,r)\,,
\eeq
\end{widetext}
which can conveniently rewritten by using the operator
\beq
\label{L_r_def0}
{\mathcal L}_{(r)}(\cdot)\equiv \frac{d^2}{dr^2}(\cdot)+\frac{\Delta'(r)}{\Delta(r)}\frac{d}{dr}(\cdot)+\left(\frac{\omega^2r^4}{\Delta^2(r)}  - \frac{L}{\Delta(r)}\right)(\cdot)\,,
\eeq
such that
\bea
{\mathcal L}_{(r)}R_{lm\omega}&=&-\frac{4\pi r^2}{\Delta(r)} \widehat T_{lm}(\omega ,r)\nonumber\\
&=& -\frac{4\pi q r^2}{\Delta(r)}   Y_{lm}^*(\frac{\pi}{2}, 0){\mathcal F}(\omega,r)\,.
\eea

This equation is solved by using the Green's function method, i.e., introducing the Green's function  
\beq
{\mathcal L}_{(r)} G_{lm\omega}(r,r')=\frac{1}{\Delta(r')}\delta(r-r')\,,\qquad \Delta(r)=r^2f(r)\,,
\eeq
where
\bea
G_{lm\omega}(r,r')&=&\frac{1}{W_{lm\omega}}[R_{\rm in}(r)R_{\rm up}(r')H(r'-r)\nonumber\\
&+& R_{\rm in}(r')R_{\rm up}(r)H(r-r')]\nonumber\\
&\equiv & \frac{1}{W_{lm\omega}}R_{\rm in}(r_<)R_{\rm up}(r_>)\,,
\eea
and $r_<$ and $r_>$ correspond to $r$ and $r'$, respectively.
$R_{\rm in}(r)$ and $R_{\rm up}(r)$ are two independent solutions of the homogeneous equation, satisfying ingoing boundary conditions at the horizon and outgoing boundary conditions at infinity, respectively. Both depend on $l,m,\omega$, which are not shown explicitly to simplify the notation. The range of the radial variable is $[r_+,\infty)$, and
\beq
W_{lm\omega}=\Delta (r) [R_{\rm in}(r)R_{\rm up}'(r)-R_{\rm up}(r)R_{\rm in}'(r)]
\eeq
is the constant Wronskian.
The solution $R_{lm\omega}(r)$ of the non-homogeneous equation is then given by
\bea
R_{lm\omega}(r)&=&-4\pi\int dr' G_{lm\omega}(r,r')r'^2 \widehat T_{lm}(\omega ,r')\nonumber\\
&=& -4\pi q Y_{lm}^*(\frac{\pi}{2}, 0)  \int dr' G_{lm\omega}(r,r')r'^2 {\mathcal F}(\omega,r')\nonumber\\
&=& -4\pi q Y_{lm}^*(\frac{\pi}{2}, 0) \int dt e^{i \omega t }  G_{lm\omega}(r,r_p(t)) f(r_p(t))\,. \nonumber\\
\eea
\begin{widetext}

Since we have both a left and a right Green function (with respect to the particle's world line) we have also a left and a right solution for $R_{lm\omega}(r)$, namely
\bea
R_{lm\omega}^-(r)
&=& -4\pi q Y_{lm}^*(\frac{\pi}{2}, 0)  \frac{R_{\rm up}(r) }{W_{lm\omega}}\int dt e^{i \omega t } R_{\rm in}(r_p(t))f(r_p(t))\,, \nonumber\\
R_{lm\omega}^+(r)
&=& -4\pi q Y_{lm}^*(\frac{\pi}{2}, 0)  \frac{R_{\rm in}(r)}{W_{lm\omega}}\int dt e^{i \omega t } R_{\rm up}(r_p(t)) f(r_p(t)) \,,
\eea
which can be used to define a left and a right field $\psi^\pm (t,r,\theta,\phi)$ by using Eq. \eqref{psi_exp0}.
However, it is easy to show that the field (but not its first derivatives, i.e., the associated  self force) is continuous along the particle's world line,   $\psi=\psi^+=\psi^-$, and one writes
\bea
\label{psi_exp0n}
\psi(t,r,\theta,\phi)&=&\sum_{lm}\int \frac{d\omega}{2\pi}e^{-i\omega t}R_{lm\omega}(r)Y_{lm}(\theta,\phi)\nonumber\\
&=& -4\pi q \sum_{lm}Y_{lm}(\theta,\phi) Y_{lm}^*(\frac{\pi}{2}, 0)\int \frac{d\omega}{2\pi}e^{-i\omega t} \frac{1}{W_{lm\omega}} \int dt' e^{i \omega t' } R_{\rm in}(r_p(t'))R_{\rm up}(r) f(r_p(t'))\,,\nonumber\\
&=& -4\pi q \sum_{lm}Y_{lm}(\theta,\phi) Y_{lm}^*(\frac{\pi}{2}, 0)R_{\rm up}(r){\mathcal N}_{lm}(t)\,,
\eea
\end{widetext}
where we have defined
\bea
\label{M_N_defs}
{\mathcal M}_{lm}(t,t')&=&\int_{-\infty}^\infty \frac{d\omega}{2\pi} \frac{e^{-i\omega (t-t')}}{W_{lm\omega}}  R_{\rm in}(r_p(t')) f(r_p(t'))\,,\nonumber\\
{\mathcal N}_{lm}(t)&=&\int_{-\infty}^\infty dt' {\mathcal M}_{lm}(t,t')\,.
\eea
One can now evaluate the  field \eqref{psi_exp0n} along the source world line obtaining a function of $t$ only,
\beq
\psi(t,r_p(t),\frac{\pi}{2},0)\,,
\eeq
as well as its  first derivatives with respect the coordinate, i.e. the self force (discontinuous along the source world line).
The latter is explicitly discussed in the next subsection.

\subsection{Scalar self force}

The self-force acting on the scalar particle is defined as
\beq
F^\alpha =q P(U)^{\alpha\beta}\partial_\beta \psi\,,
\eeq
where $P(U)=g+U\otimes U$ projects orthogonally to $U$, the particle's four velocity. 
Explicitly,
\bea
F_\alpha  
&=& q \left(\partial_\alpha \psi + U_\alpha U^\beta \partial_\beta \psi \right)\,,
\eea
that is
\bea
F_t   &=&q\sqrt{\frac{2M}{r}}\left(-\sqrt{\frac{2M}{r}}\frac{1}{f}\partial_t \psi+ \partial_r \psi\right)\,,\nonumber\\
F_r   &=& 
-\frac{U^t}{U^r}F_t\,,\quad
F_\theta = q  \partial_\theta \psi\,, \quad
F_\phi   =q  \partial_\phi \psi \,,
\eea
where we recall the expressions of the (covariant, contravariant) components of the four velocity 
\bea
U_t&=&-1\,,\qquad U_r=-\frac{1}{f}\sqrt{\frac{2M}{r}} \,,\nonumber\\
U^t&=&\frac{1}{f}\,,\qquad U^r=-\sqrt{\frac{2M}{r}}\,.
\eea
The proportionality between $F_t$ and $F_r$ follows from the orthogonality condition $F\cdot U=0$.
The components of the force should be evaluated along the particle's orbit, and then become functions of the proper time (or any other parameter used to parametrize the orbit, like the coordinate time or the radial position itself). 
 
\subsection{Radiation losses: formal expressions}

The expressions for the radiated energy and angular momentum can be derived from the associated energy-momentum tensor which for a 
massless complex scalar field is given by
\beq
8\pi T^{\rm scal}_{\mu\nu}=\psi^*_{,\mu}\psi_{,\nu}+\psi_{,\mu}\psi^*_{,\nu}-  g_{\mu\nu} \psi^*_{,\alpha}\psi^{,\alpha} \,,
\eeq
so that
\beq
\frac{d^2E}{dtd\Omega}=\lim_{r\to \infty} (r^2 T^{\rm scal}{}^r{}_t)\,.
\eeq
One finds
\beq
8\pi T^{\rm scal}_{rt}=\psi^*_{,r}\psi_{,t}+\psi_{,r}\psi^*_{,t} \,,
\eeq
implying
\beq
8\pi T^{\rm scal}{}^r{}_{t}=f(\psi^*_{,r}\psi_{,t}+\psi_{,r}\psi^*_{,t})=f \psi^*_{,r}\psi_{,t}+{\rm c.c.}\,.
\eeq
Let us use the labels \lq\lq $l,m,\omega$" for the $t$ derivative and $l'm',\omega'$" for the $r$ derivative of $\psi$, namely
\bea
\psi&=& \sum_{lm}Y_{lm}\int \frac{d\omega}{2\pi}e^{-i\omega t}R_{lm\omega}\,,\nonumber\\
\psi_{,t}&=& \sum_{lm}Y_{lm}\int \frac{d\omega}{2\pi}(-i\omega) e^{-i\omega t}R_{lm\omega}\,,\nonumber\\
\psi^*_{,r}&=& \sum_{l'm'}Y_{l'm'}^*\int \frac{d\omega'}{2\pi}e^{i\omega' t}\frac{d}{dr}R_{l'm'\omega'}^*\,,
\eea 
and
\begin{widetext}
\bea
r^2 T^{\rm scal}{}^r{}_{t}&=&\frac{r^2 f}{8\pi} (\psi^*_{,r}\psi_{,t}+\psi_{,r}\psi^*_{,t})\nonumber\\
&=& \frac{r^2 f}{8\pi} \sum_{lm,l'm'}\int \frac{d\omega}{2\pi} \frac{d\omega'}{2\pi}Y_{lm}(\theta,\phi)Y_{l'm'}^*(\theta,\phi)  \left[(-i\omega) e^{-i(\omega-\omega')t} R_{lm\omega}(r)\frac{d}{dr}R^*_{l'm'\omega'}(r)\right.\nonumber\\
&+& \left. (i\omega')e^{i(\omega-\omega')t} R^*_{l'm'\omega'}(r)\frac{d}{dr}R_{l m \omega }(r)\right]\,.
\eea
Integrating over the sphere and recalling the orthogonality relation \cite{NIST} (see Eq. 14.30.8 there) 
\beq
\int \sin \theta d\theta d\phi \,  Y_{lm}^*(\theta,\phi)Y_{l'm'}(\theta,\phi)=\delta_{ll'}\delta_{m m'}\,,
\eeq
we find
\bea
\frac{dE}{dt}&=&\int \sin \theta d\theta d\phi r^2 T^{\rm scal}{}^r{}_{t}\nonumber\\
&{=}&\left. \frac{r^2 f}{8\pi} \sum_{lm}\int \frac{d\omega}{2\pi} \frac{d\omega'}{2\pi}\left[({-}i\omega) e^{{-}i(\omega{-}\omega')t} R_{lm\omega}(r)\frac{d}{dr}R^*_{lm\omega'}(r){+}(i\omega')e^{i(\omega{-}\omega')t} R^*_{lm\omega'}(r)\frac{d}{dr}R_{l m \omega }(r)\right]\right|_{r=r_p(t)}\,.\nn\\
\eea
In the second integral we can exchange $\omega$ and $\omega'$, obtaining
\bea
\frac{dE}{dt}&=& \frac{r^2 f}{8\pi} \sum_{lm}\int \frac{d\omega}{2\pi} \frac{d\omega'}{2\pi}(-i\omega)e^{-i(\omega-\omega')t} \left[ R_{lm\omega}(r)\frac{d}{dr}R^*_{lm\omega'}(r)
-R^*_{lm\omega}(r)\frac{d}{dr}R_{l m \omega' }(r)\right]\big|_{r=r_p(t)}\nonumber\\
&=& \frac{r^2 f}{8\pi} \sum_{lm}\int \frac{d\omega}{2\pi} \frac{d\omega'}{2\pi}(-i\omega)e^{-i(\omega-\omega')t} 2i {\rm Im}\left(R_{lm\omega}(r)\frac{d}{dr}R^*_{lm\omega'}(r)\right)\big|_{r=r_p(t)}\nonumber\\
&=& \frac{r^2 f}{4\pi} \sum_{lm}\int \frac{d\omega}{2\pi} \frac{d\omega'}{2\pi} \omega e^{-i(\omega-\omega')t} {\mathcal J}_{lm\omega \omega'}(r)\big|_{r=r_p(t)}\,,
\eea
\end{widetext} 
where we have defined
\beq
{\mathcal J}_{lm\omega \omega'}(r)=
{\rm Im}\left(R_{lm\omega}(r)\frac{d}{dr}R^*_{lm\omega'}(r)\right)\,,
\eeq
and in which one has to use  $R_{lm\omega}(r)=R_{lm\omega}^+(r)$, namely,   
\beq
R_{lm\omega}^+(r)
= -4\pi q Y_{lm}^*(\frac{\pi}{2}, 0)  \frac{R_{\rm in}(r)}{W_{lm\omega}}\int dt e^{i \omega t }R_{\rm up}(r)f(r)|_{_{r=r_p(t)}}\,. 
\eeq
because of the evaluation at $t\to +\infty$.
Therefore,
\bea
\Delta E&=& \int dt \frac{dE}{dt}\nonumber\\
&=& -\int_{2M}^\infty \frac{dt}{dr} dr  \frac{r^2 f}{4\pi} \times \nonumber\\
&& \sum_{lm}\int \frac{d\omega}{2\pi} \frac{d\omega'}{2\pi} \omega e^{-i(\omega-\omega')t} {\mathcal J}_{lm\omega \omega'}(r)\,.
\eea
The final expression for the radiated energy 
\beq
\frac{dE}{dt}=\lim_{r\rightarrow\infty}\int \sin\theta d\theta d\phi r^2 {T^{{\rm scal}\,r}}_t
\eeq
follows straightforwardly 
\bea
\frac{dE}{dt} 
&=&\lim_{r\rightarrow \infty}\frac{r^2f(r)}{8\pi}\sum_{l m}\int \frac{d\omega}{2\pi}\frac{d\omega'}{2\pi}\Bigg[\nonumber\\
&&(-i\omega)e^{-i(\omega-\omega')t}R_{l m \omega}(r)\frac{d}{dr}R^*_{l m \omega'}(r)\nonumber\\
&+&(i\omega')e^{i(\omega-\omega')t}R^*_{l m\omega'}(r)\frac{d}{dr}R_{l m \omega}(r)\Bigg]\,.
\eea
As already stated, here, because of  the $r\rightarrow\infty$ limit, we must consider only the up-part of the $R_{l m \omega}(r)$ solution.

In this section, our main result in the scalar field case, including contributions from $l=0,1,2$ only and PN expanding up to the $O(\eta^8)$ fractional accuracy, is the following
\begin{widetext}
\bea
\label{E_rad_scal}
E_{\rm rad}&=& \int_{-\infty}^\infty dt \frac{dE}{dt}\nonumber\\
&=&  \frac{q^2}{M}\Big[-\frac{\eta ^3}{90}-\frac{739 \eta ^5}{1512}+\left(\frac{2133 \sqrt{3}}{14000}-\frac{49}{360}\right) \pi  \eta ^6+\frac{2541439 \eta ^7}{1360800}+\left(\frac{8}{3}-\frac{25970213}{2205000 \sqrt{3}}\right) \pi  \eta ^8\Big]+O\left(\eta^9\right)\,.\qquad
\eea
\end{widetext}
In the next section, instead, we will  consider  the more interesting case of the gravitational perturbations.

\section{Gravitational perturbations: analytic computations of the radiated energy at the 2PN fractional accuracy}

It is instructive to start quickly by the simplest case of the leading Newtonian order (actually, a warming-up exercise; see e.g. Refs \cite{Maggiore:2007ulw,Maggiore:2018sht}), passing then to a Post-Newtonian (PN) computation which will be used later to test the first few terms of the genuine self force computation we are going to develop in detail below.  

\subsection{The leading Newtonian order}

The computation of the energy emitted by a test particle of mass $m$
radially infalling into a Schwarzschild black hole requires several assumptions, already at the Newtonian level. 
First, the gravitational wave (GW) production is computed in the linearized theory, without including backreactions on the Schwarzschild metric. Second, the Newtonian equations of motion are used instead of the general relativistic ones, 
\beq
\label{moto}
\frac{1}{2}m \left(\frac{dr_p}{dt}\right)^2 -\frac{G m M}{r_p(t)}=0\,,
\eeq
implying
\be\label{NEM}
\frac{dr_p}{dt}=-\sqrt{\frac{2GM}{r_p(t)}}\,.
\ee
The third assumption requires  that most of the radiation is emitted when the particle is non relativistic, and therefore the quadrupole formula is still valid (namely, until the particle enters the strong field region, i.e., approaches the black hole horizon). Currently, strong field computations can be performed only numerically.

In fact, in the last part of the radial fall orbit all these assumptions break down. 
If one evaluates Eq. \eqref{NEM} at the horizon $r=2GM/c^2$ one obtains $\dot{r}(t)=-c$, obviously beyond the slow-motion approximation. Thus, one can only compute the radiation emitted from $r=+\infty$ until a certain radius $r=R>2M$ (to be fixed later).  

The solution of the Newtonian equation of motion (such that at $t=t_0$ the particle is at the  radius $r(t_0)$) is given by
\beq\label{NEQ2}
r^{3/2}(t)-r^{3/2}(t_0)=\frac{3}{2} \sqrt{2GM}(t_0-t)\,,
\eeq
where the boundary conditions are chosen so that there exists a maximal value of the time at which the particle reaches $R$,  the boundary of the strong field region
\beq
\lim_{t\to -\infty }r(t)=+\infty\,,\qquad r(t_{\rm max})=R\,.
\eeq
See Ref. \cite{Maggiore:2007ulw} for a detailed and pedagogical introduction to this problem.
Let us define
\be
\frac{3}{2}\sqrt{2GM} \bar{t}\equiv\frac{3}{2}\sqrt{2GM} t_0+r^{3/2}(t_0)\,,
\ee
so that Eq. \eqref{NEQ2} becomes
\beq
r^{3/2}(t)=\frac{3}{2}\sqrt{2GM}(\bar{t}-t)\,,
\eeq
implying that $r(\bar t)=0$. Equivalently,
\beq
r(t)=\left(\frac{3}{2}\sqrt{2GM}\right)^{2/3}(\bar{t}-t)^{2/3} \equiv \alpha (\bar{t}-t)^{2/3}\,,
\eeq
with
\beq
\alpha =\left( \frac92 GM\right)^{1/3}\,,
\eeq
already defined in Eq. \eqref{alpha_def}.

We find $\lim_{t\to -\infty}r(t)=+\infty$, while the minimum value $r(t)=R$ is reached at 
\be
t_{\rm max}=\bar{t}-\frac{2R^{3/2}}{3 \sqrt{2GM}}\,.
\ee
Introducing a shifted temporal coordinate $\bar{t}-t=\tau$, the solution of the equations of motion reads
\beq
r(\tau)=\alpha \tau^{2/3}=\left(\frac{3}{2}\sqrt{2GM} \tau\right)^{2/3}\,.
\eeq
The radiated energy at this (Newtonian) level can be computed via the well-known quadrupole formula
\beq
E_{\rm rad}=\frac{G}{5 c^5}\int_{-\infty}^{t_{\rm max}} dt I_{ij}^{(3)}(t)I_{ij}^{(3)}(t)\,,
\eeq 
where $I_{ij}(t)= \mu x^{\langle i} x^{j\rangle}$ (while $m$ in Eq. \eqref{moto} is the mass of the perturbing particle, $\mu$ is the reduced mass of the two-body system) and, we recall, the  process should be limited to the interval $t\in (-\infty,t_{\rm max})$.

In our case (equatorial motion, radial fall along the $x$-axis, $\phi(t)=0$)
\beq
I_{ij}(t)=\frac13 \mu r^2 q_{ij}\,,\qquad q_{ij}={\rm diag}[2,-1,-1]\,,
\eeq
with $q_{ij}$ symmetric and tracefree and such that
\beq
q_{ij}q_{ij}=\delta^{ij}(q^2)_{ij}=6\,,
\eeq
and 
\beq
\frac{G}{5 c^5}I_{ij}^{(3)}(t)I_{ij}^{(3)}(t)=\frac{2G}{15 c^5}\mu^2 \frac{(2GM)^3}{r^5(t)}\,.
\eeq
Consequently,
\bea
E_{\rm rad}&=& \frac{2G(2GM)^3}{15 c^5}\mu^2  \int_{-\infty}^{t_{\rm max}}  \frac{dt}{r^5(t)}\nonumber\\
&=& Mc^2 \nu^2  \frac{ 2^{\frac{17}3}3^{-\frac{10}3}  \left(\frac{G M}{c^3 \tau_{\rm min}} \right)^{\frac{7}{3}}}{35 }  \,,
\eea
with $\tau_{\rm min}=(R/\alpha)^{3/2}$ and $\frac{G M}{c^3 \tau_{\rm min}}$ dimensionless. 
In the Fourier space
\bea
\label{FT_I_ij}
(-i\omega)^3 \hat I_{ij}(\omega)&=& \int_{-\infty}^\infty dt e^{i\omega t} I_{ij}^{(3)}(t)\nonumber\\
&=&\frac13 \mu (2GM)^{3/2} q_{ij} \int_{-\infty}^\infty dt \frac{e^{i\omega t}}{r^{5/2}(t)} \nonumber\\
&=& \mu \frac13 \frac{(2GM)^{3/2}}{\alpha^{5/2}} e^{i\omega \bar t}J(\omega) q_{ij}\,,  
\eea
where
\bea
\label{J_def}
J(\omega)&=&\int_{-\infty}^\infty d\tau  \frac{e^{-i\omega \tau}}{\tau^{5/3}} 
\nonumber\\
&=&  \sqrt{3}  e^{\frac16 i\pi}  |\omega|^{2/3} \Gamma\left(-\frac{2}{3}\right)   \theta(\omega)\,,
\eea
with  
$\theta(\omega)$ the Heaviside step function.

In the Fourier space we have then
\bea
E_{\rm rad}&=& \frac{G}{5c^5}\int \frac{d\omega}{2\pi}  \omega^6 |I(\omega)|^2\nonumber\\
&=& \int d\omega \frac{dE}{d\omega} 
\eea
with
\beq
\frac{dE}{d\omega} =\frac{1}{2\pi}\frac{G}{5c^5}  \omega^6 |I(\omega)|^2\,.
\eeq
We find
\bea
\frac{dE}{d\omega} &=&  \frac{1}{2\pi} \frac{G}{5c^5}  \mu^2 \frac{(2GM)^{3 }}{\alpha^{5}} \frac{6}{9}  |J(\omega)|^2  \nonumber\\
 &=&  \frac{G}{ 5\pi c^5}   \mu^2 \frac{(2GM)^{3 }}{\alpha^{5}} |\omega|^{4/3}\Gamma^2\left(-\frac{2}{3}\right)\theta(\omega)\,.
\eea
Summarizing, for $\omega>0$ we find
\bea
\frac{dE}{d\omega} 
 &=&  {\mathcal E}  \omega ^{4/3} \,,
\eea
where  
\bea
{\mathcal E} &=& \frac{ G}{ 5\pi c^5}   \mu^2 \frac{(2GM)^{3 }}{\alpha^{5}}  \Gamma^2\left(- \frac{2}{3}\right)\nonumber\\
&=&  \frac{\nu^2 M c^2}{ 5\pi  } \left(\frac{2}{3} \right)^{7/3}   \left(\frac{2GM}{c^3}\right)^{7/3}  \Gamma^2\left(-\frac{2}{3}\right)\,.
\eea
This implies that in the Fourier space the (integrated) energy loss due to quadrupolar emission ($l=2$) is divergent and one has to truncate the result at a certain value of $\omega=\omega_{\rm max}$ corresponding to the entrance of the particle in the strong field region marked by $r=R$. As mentioned above, at this point the PN approximation breaks down and one must resort to other approximation techniques. We therefore will limit to display energy densities (in the time domain and in the Fourier domain) which are not affeted by this problem. Let us note, in passing, that a Post-Minkowskian-type treatment of the problem can be of interest but a complete technology for this to be applied to black hole perturbations is not available yet (unless one considers scattering processes where, differently, approaches of the type of effective field theory and \lq\lq amplitudes" already play a role, see e.g., \cite{Kosower:2018adc,Bern:2019nnu,Bern:2021dqo,Dlapa:2021npj,Bern:2025zno}).

\subsection{The  complete 2PN result}

The harmonic coordinate 2PN conserved energy (per unit of reduced mass) reads
\bea
\mathcal{E}&=&\mathcal{E}_{\rm N}+\eta^2 \mathcal{E}_{\rm 1PN}+\eta^4 \mathcal{E}_{\rm 2PN}\,,
\eea
where
\bea
\mathcal{E}_{\rm N}&=&\frac{1}{2}v^2-\frac{G M}{r}\,,\nn\\
\mathcal{E}_{\rm 1PN}&=&\frac38(1-3\nu)v^4+\frac12(3+\nu)v^2 \frac{GM}{r}+\frac12 \nu \dot r^2 \frac{GM}{r}\nonumber\\ 
&+&\frac12 \left(\frac{GM}{r}\right)^2\,,\nn  \\
\mathcal{E}_{\rm 2PN}&=&\frac{5(1-7\nu+13\nu^2) }{16}v^6 + \frac{21-23\nu-27\nu^2}{8}v^4 \frac{GM}{r}\nonumber\\
&+& \frac{\nu(1-15\nu)}{4}\dot r^2 v^2 \frac{GM}{r}-\frac{3\nu (1-3\nu)}{8}\dot r^4\frac{GM}{r} \nonumber\\
&-&\frac{2+15\nu}{4} \left(\frac{GM}{r}\right)^3
+\frac{14-55\nu+4\nu^2}{8}\left(\frac{GM}{r}\right)^2 v^2 \nonumber\\
&+&\frac{4+69\nu+12\nu^2}{8}\left(\frac{GM}{r}\right)^2 \dot r^2
\,,
\eea
where 
\be
v=\sqrt{\dot{r}^2+r^2\dot{\phi}^2}\,.
\ee
For $\dot{\phi}=0=J$, one finds $v \to |\dot r|$, 
and then 
$\mathcal{E}=\mathcal{E}(r,\dot r)$.
Assuming ${\mathcal E}=0$ (releasing from rest) this equation defines $\dot r $ as a function of $r$, i.e., a (perturbative) differential equation which can be easily solved 
to get the harmonic coordinates and 2PN accurate orbit $r=r_p(t)$,
\bea
\label{rp}
r_p(t)&=&  \frac{ 3^{2/3} (G M)^{1/3}   (-t)^{2/3} }{ 2^{1/3}}+
\frac{5}{2} G M (\nu-2) \eta^2\nonumber\\ 
&+& 
 \frac{ (G M)^{5/3} (48 - 19 \nu + 5 \nu^2) }{ 
  2\cdot 6^{2/3} (-t)^{2/3} }\eta^4 \nonumber\\
&+& O(\eta^6)\,.
\eea
Let us recall that the particle is released from rest at radial infinity (at $t\to -\infty$), corresponding to a vanishing binding energy ${\mathcal E}$, besides the obvious vanishing of angular momentum due to the radial fall.  Furthermore, as a consequence of our choice of the parametrization of the orbit, the radial velocity $\dot{r}$ is always negative.

The energy flux at the 2PN level of accuracy is given by
\bea
\label{Fexp}
\mathcal{F}&=&\mathcal{F}_{\rm N}+\eta^2\mathcal{F}_{\rm 1PN}+\eta^4\mathcal{F}_{\rm 2PN}\,,
\eea
where the explicit expressions of $\mathcal{F}_{\rm N}$, $\mathcal{F}_{\rm 1PN}$ and $\mathcal{F}_{\rm 2PN}$ for a generic orbit can be found in previous literature, e.g. \cite{Arun:2007sg} (see Eqs. 5.2 there and see also \cite{Blanchet:2013haa} for a review), and are included here only for completeness. Specializing them to the radial fall ($v=|\dot r|=-\dot r$) each PN contribution to the flux reduces to a function of $r$ and $\dot r$. Explicitly,
\bea
\label{Fordrad}
\mathcal{F}_{\rm N}&=& \frac{32G^3M^4\nu^2}{5c^5r^4} \frac{1}{12}\dot r^2\,,\nonumber\\
\mathcal{F}_{\rm 1PN}&=&\frac{32G^3M^4\nu^2}{5c^7r^4} \Big(\frac{(1-4 \nu ) G^2 M^2}{21
r^2}\nn\\
&+&\frac{ (5 \nu +27) GM
\dot r^2}{42 r}+\frac{1}{42}
(9 \nu -16)\dot r^4\Big)\,, \nonumber \\
\mathcal{F}_{\rm 2PN}&{=}& \frac{32G^3M^4\nu^2}{5c^9r^4} \Bigg[
{-}\frac{ \left(56 \nu ^2{-}1026
   \nu {+}253\right) G^3M^3}{378
   r^3}\nn\\
   &-&\frac{ \left(5880
   \nu ^2-26217 \nu +18742\right)
  G^2M^2 \dot r^2}{4536
  r^2}\nonumber\\
&+&\frac{ \left(1016 \nu^2+65 \nu -1965\right) GM
\dot r^4}{756
r}\nn\\
&+&\frac{1}{252} \left(-67
\nu ^2-25 \nu +44\right)
\dot r^6
\Bigg]\,.
\eea

Inserting Eq. \eqref{rp} in Eq. \eqref{Fexp} and using Eqs. \eqref{Fordrad}, we obtain the final 2PN accurate result 
\bea
\frac{\mathcal{F}}{\eta^5\nu^2}&=&{\mathcal F}^{\nu^0}+{\mathcal F}^{\nu^1}\nu +{\mathcal F}^{\nu^2}\nu^2\,,
\eea
where
\bea
{\mathcal F}^{\nu^0}&=& \frac{32\cdot 2^{2/3}}{405\cdot 3^{1/3}T^{10/3}}{+}\frac{2816\eta^2}{2835T^4}{+}\frac{1076224\cdot 2^{1/3}\eta^4}{229635\cdot 3^{2/3}T^{14/3}}\nonumber\\
{\mathcal F}^{\nu^1}&=&-\frac{2048 \eta
   ^2}{8505 T^4} -\frac{25088 \cdot 2^{1/3} \eta
   ^4}{10935\cdot 3^{2/3}
   T^{14/3}}\nonumber\\
{\mathcal F}^{\nu^2}&=& \frac{7936 \cdot 2^{1/3} \eta
   ^4}{25515 \cdot 3^{2/3} T^{14/3}}\,.
\eea
where here we set  $G=1$ and $T=t/M$, for convenience.
In the self force limit ($\nu \to 0$) $\frac{\mathcal{F}}{\eta^5\nu^2}$ reduces to   ${\mathcal F}^{\nu^0}$ and will be used to check the self force computations in the next section.

In order to have additional checks one can also compute  the tail part of the energy flux (at the leading order, see e.g. Eq. (3.4) in Ref. \cite{Bini:2021qvf}),
given by
\bea
E_{\rm rad, tail}&=& -\frac{4G^2M}{5}\eta^8 \int \frac{d\omega}{2\pi}\omega^8 |\hat I_{ij}(\omega)|^2 A_1(\omega)\,,\qquad
\eea
with 
\bea
A_1(\omega)&=&\int_0^\infty d\tau e^{i\omega \tau}\ln \left( \frac{\tau}{C_{I_2}}\right)\nonumber\\
&=& -\frac{\pi}{2|\omega|}-\frac{i}{\omega}\ln (C_{I_2}|\omega|e^\gamma)\,, 
\eea
and $C_{I_2}$ a scale associated with the mass quadrupole moment 
\beq
C_{I_2}=r_0\frac{2e^{-11/12}}{c}\,,
\eeq
and $\gamma$ is the Euler-Mascheroni constant. Then
\bea
E_{\rm rad, tail}&=& -\frac{4G^2M}{5}\pi \eta^8 \int_0^\infty \frac{d\omega}{2\pi}\omega^7 |\hat I_{ij}(\omega)|^2\,,\qquad\nonumber\\
&=& -\frac{27}{10}C^2 \omega_{\rm max}^{10/3}\Gamma^2\left(\frac73\right)\,,
\eea
where 
\beq
C=\frac{3^{1/3}G^{2/3}M^{5/3}\nu }{2^{2/3}}\,,
\eeq
and the integration over the frequencies has been truncated at $\omega=\omega_{\rm max}$ corresponding to the entrance of the particle in the strong field region.

\section{Gravitational wave equation and Teukolsky formalism}

In this section we follow Ref. \cite{Sasaki:2003xr} for notation and conventions, i.e. the main reference review on this topic.
In fact it includes the procedure to determine the so called Mano-Suzuki-Takasugi (MST) \cite{Mano:1996vt}
type solution of the homogeneous wave equation  (Teukolsky equation, \cite{Teukolsky:1974yv}) which incorporate the correct boundary conditions both at the  horizon and at spatial infinity.
Let us denote these homogeneous solutions   as $R^{\rm in}_{l m \omega}(r)$ (ingoing at the horizon) and $R^{\rm up}_{l m \omega}(r)$ (outgoing at the spatial infinity). 
For radially infalling particles, the source term  reduces to
\bea
&&T_{l m \omega}(r)= 8\pi\mu \int_{-\infty}^{\infty}dt\int d\theta e^{i\omega t}\nonumber\\
&\times&\Bigg[{-}\frac{1}{2}L_1^\dagger\Bigg\{ L_2^\dagger\left( {}_{-2}Y_{lm} \right)\Bigg\}C_{nn}r^4\delta(r{-}r(t))\delta(\theta-\theta(t))\Bigg]\,,\nonumber\\
&&C_{nn}=\frac{(E+\frac{dr}{d\tau})^2}{4r^2\dot{t}}\,,\nonumber\\
&&L_s^\dagger=\partial_\theta-\frac{m}{\sin\theta}+s \cot\theta\,,
\eea
where ${}_{-2}Y_{l m}(\theta,\phi)$ is the $s=-2$ spin weighted spherical harmonic obeying the equation
\bea
&&0=\frac{1}{\sin(\theta)}\partial_\theta(\sin(\theta)\partial_\theta {}_{-2}Y_{l m})\nonumber\\
&+&\Bigg[-\frac{(m-2\cos(\theta))^2}{\sin^2\theta}-2+(l+2)(l-1)\Bigg]{}_{-2} Y_{l m}\,.\nonumber\\
\eea

The stress energy tensor specialized to equatorial motion reads
\bea
T_{l m \omega}(r)&=&  \int_{-\infty}^{\infty}dt e^{i\omega t}\delta(r-r(t))\frac{(E+\frac{dr}{d\tau})^2r^2}{\dot{t}}K_{lm}\,,\qquad
\eea
where $\tau$ denotes the proper time of the particle's orbit, $S={}_{-2}Y_{l m}(\frac{\pi}{2},0)$, $S'=\partial_\theta \left({}_{-2}Y_{l m}(\theta,\phi)\right)|_{(\frac{\pi}{2},0)}$
and
\bea
K_{lm}&=& 2\pi \mu \Bigg[\frac{1}{2}(l^2+l-2-2m^2)S+mS'\Bigg]\nonumber\\
&=& -(2\pi)\frac{\mu}{2} L_1^\dagger  L_2^\dagger\left( {}_{-2}Y_{lm} \right)\nonumber\\
&\equiv & 2\pi \mu\bar K_{lm}\,.
\eea
For example,  
\bea
\bar K_{2,-2} &=&\bar K_{2,2} =-\frac34 \sqrt{\frac{5}{\pi}}\,,\nonumber\\
\bar K_{2,-1} &=&\bar K_{2,1} =0\,,\nonumber\\ 
\bar K_{2,0} &=& \frac12 \sqrt{\frac{15}{2\pi}}\,.
\eea
Using the geodesic equations and the fact that the particle starts its motion from rest, we have
\bea\label{Tlmomega}
T_{l m \omega}(r)&=& K_{lm} \int_{-\infty}^{\infty}dt e^{i\omega t}\delta(r-r(t))\left(1+\frac{dr}{d\tau}\right)^2\Delta(r)\,,\nonumber\\
\eea
where $\Delta(r)=r(r-2M)$\,.

Recalling Eqs. \eqref{dt_dtau}, \eqref{dr_dt} and \eqref{dr_dt2}
for a particle released from rest at infinity 
we find
\bea\label{Tlmomega2}
T_{l m \omega}(r)&=& K_{lm}\int_{-\infty}^{\infty}dt e^{i\omega t}\delta(r-r(t))(1+\frac{dr}{d\tau})^2r(r-2M)\nonumber\\
&=& K_{lm}\int_{-\infty}^{\infty}dt e^{i\omega t}\delta(r{-}r(t))(\sqrt{r}{-}\sqrt{2M})^3\times \nonumber\\
&& (\sqrt{r}{+}\sqrt{2M})\,,
\eea
where we have used the relation $\frac{dr}{d\tau}=-\sqrt{2M/r}$ and
\beq
rf=r-2M=(\sqrt{r}-\sqrt{2M})(\sqrt{r}+\sqrt{2M})\,.
\eeq

Finally, we will use  the \lq\lq tortoise coordinate" \cite{Misner:1973prb}
\beq
r_*=r+2M \ln \left(\frac{r}{2M}-1\right)\,,
\eeq
as well as 
the asymptotic relations valid for $r\to \infty$
\bea
R^{\rm in}_{l m \omega}(r)&=& r^3 B^{\rm ref}_{l m\omega}e^{i\omega r_*}+\frac1r B^{\rm inc}_{l m\omega}e^{-i\omega r_*}\,,\nonumber\\
R^{\rm up}_{l m \omega}(r)&=& C^{\rm trans}_{l m \omega} r^3 e^{i\omega r_*}\,,
\eea
implying
\beq
W_{l m \omega}=2i\omega C^{\rm trans}_{l m\omega} B^{\rm inc}_{l m\omega}\,.
\eeq
Moreover $C^{\rm trans}_{l m\omega}$ can be given a closed form expression
\bea
C^{\rm trans}_{l m\omega}&=&\omega^3  e^{i \epsilon \ln \epsilon }A_-^\nu\nonumber\\
A_-^\nu &=& 2^{1+i\epsilon}e^{-i\frac{\pi}{2}(\nu-1)}e^{-\frac{\pi}{2}\epsilon}\sum_{n=-\infty}^\infty (-1)^n a_n \frac{(\nu-1-i\epsilon)_n}{(\nu+3+i\epsilon)_n}.\nonumber\\
\eea
The amplitude of the wave at infinity then reads
\bea
R_{l m \omega}(r)&\approx &\frac{C^{\rm trans}_{l m \omega} r^3 e^{i\omega r_*}}{W_{l m \omega }}\int _{2M}^\infty dr' \frac{R^{\rm in}_{l m \omega}(r')T_{l m\omega}(r')}{\Delta^2(r')}\nonumber\\
&\equiv& r^3 e^{i\omega r_*} \widehat Z_{l m \omega}\,,
\eea
where
\bea\label{hatZlm}
\widehat Z_{l m \omega}&=&\frac{C^{\rm trans}_{l m \omega}}{W_{l m \omega}}\int_{2M}^\infty dr\frac{R^{\rm in}_{l m \omega}(r)T_{l m \omega}(r)}{\Delta(r)^2}\nonumber\\
&=&
\frac{C_{l m \omega}^{\rm trans}}{W_{l m \omega}}K_{lm}\int_{-\infty}^{\infty}dt e^{i\omega t}R^{\rm in}_{l m \omega}(r(t))\times \nonumber\\
&&\frac{(\sqrt{r(t)}{-}\sqrt{2M})}{r(t)^2(\sqrt{r(t)}{+}\sqrt{2M})}\,.
\eea
We have used here several results from Fourier domain integrals which we conveniently discuss in the next subsection.

\subsection{The needed Fourier domain results}
Most of the difficulties associated with the present computation are related to the way in which certain basic \lq\lq master integrals" are evaluated.
Notice that  with our choice of the coordinate representation of the orbit  the falling particle uses negative values of the coordinate time, starting from $t\to -\infty$ (when it is released from $r\to +\infty$).
They are the following
\bea
&&I_s(\omega) = \int_{-\infty}^\infty dt e^{i\omega t} t^s\nn\\
&=&-2e^{\frac{i\pi s}{2}} |\omega|^{-s-1} \Gamma(1+s) \sin (\pi s)\theta(-\omega)\nonumber\\
&&I_s^{\rm ln }(\omega) =\frac{d}{ds}I_s(\omega)= \int_{-\infty}^\infty dt e^{i\omega t} t^s \ln t\nonumber\\
&=& -e^{\frac{i\pi s}{2}}  |\omega|^{-s-1}  \Gamma(1+s)\theta(-\omega)\times \nonumber\\
&&\left[2\pi \cos(\pi s) {+} \sin(\pi s) \left(i\pi {-}\ln (\omega^2){+}2\Psi^{(0)}(1{+}s) \right) \right]\,,\qquad\nonumber\\
\eea
which is valid for any real non integer $s$.

When $s$ is a non negative integer instead we have
\bea
I_n(\omega) &=& \frac{2\pi}{i^n}\delta^{(n)}(\omega)\,,\nonumber\\
I_n^{\rm ln}(\omega)&=&\frac{2 i^n n!\pi \theta(-\omega)}{\omega^{n+1}}\nonumber\\
&+& \pi (-i)^{n-1}(\pi+2i\gamma)\delta^{(n)}(\omega)\,.
\eea
Moreover, we find
\bea
J_s(t)&=& \int_{-\infty}^\infty \frac{d\omega}{2\pi} e^{-i \omega t}\omega^s \theta(-\omega)\nonumber\\
&=& \frac{i^{1+s} (-1)^s  \Gamma (s+1)}{2\pi}t^{-s-1}\,,
\eea
with $s\not \in \mathbb{N}$ and
\bea
J_s^{\rm ln }(t)&=&=\frac{d}{ds}J_s(t)\nonumber\\
&=& \int_{-\infty}^\infty \frac{d\omega}{2\pi} e^{-i \omega t}\omega^s \theta(-\omega)\ln\omega\nonumber\\
&=&-\frac{1}{4\pi}(-i)^s t^{-s-1} \Gamma(1+s) \times \nonumber\\
&& \left( 3\pi +i \ln t^2-2i \Psi^{(0)}(s+1)\right)\,,
\eea
where
\beq
\Psi^{(k)}(z)=\frac{d^k}{dz^k}\Psi(z)\,,\qquad \Psi(z)\equiv\Psi^{(0)}(z) =\frac{d}{dz}\ln \left(\Gamma(z) \right)\,,
\eeq
is the Polygamma function. $J_s(t)$ and $J_s^{\rm ln }(t)$ appear in our computations only with non integer $s$. Finally, higher powers of logs
\beq
I_{s,p}^{\rm ln }(\omega) = \int_{-\infty}^\infty dt e^{i\omega t} t^s \ln^p t\,,
\eeq 
can be obtained from $I_{s,0}^{\rm ln }(\omega)\equiv I_{s}^{\rm ln }(\omega)$ by taking further derivatives with respect to $s$.

From a technical point of view we will proceed here by integrating in $r$ first (trivialized by the presence of the Dirac delta function $\delta (r-r_p(t))$, then over time, and finally over the frequencies. We have checked that interchanging these operations is allowed and leads to the same results. Furthermore, there are possible different ways to write the formal solution of the integrals above. We have checked that with specific values of $s$  these ways all lead to the same results, i.e. the ones written above.

\subsection{The 1SF waveform and the radiated energy}

We can pass now to the waveform $h$. Let us recall that $\dot{h}$ is obtained by integrating  $\psi_{4}$,
\beq
\psi_4(r\to \infty)\sim -\frac{1}{2}(\ddot{h}_+-i\ddot{h}_\times)=-\frac{1}{2}\ddot{h}\,.
\eeq
One finds
\be
\dot{h}=\frac{2}{r}\sum_{l m}\int\frac{d\omega}{2\pi}\frac{\widehat Z_{l m \omega}}{i\omega}e^{-i\omega t}{}_{-2}Y_{l m}(\theta,\phi)\,,
\ee
and the $O(\mu^2)$ energy flux at infinity is given by
\be
\frac{dE_{\rm rad}}{dt}=\frac{r^2}{16\pi}\int d\Omega|\dot{h}|^2\,,
\ee
since $h \propto \mu/M$ is known only at first order in the ratio $\mu/M$.
We find
\bea
\label{eq_dE_rad}
\frac{dE_{\rm rad}}{dt}
&=&\frac{1}{4\pi}\sum_{l ,m}\int \frac{d\omega}{2\pi}\frac{d\omega'}{2\pi}e^{-i(\omega-\omega')t}\frac{\widehat Z_{l m \omega}}{i\omega}\frac{\widehat Z_{l m \omega'}^*}{-i\omega'}\,,\nonumber\\
\eea
having used the relation valid for spin-weighted spherical harmonics
\beq
\int d\Omega \, {}_sY_{l m}(\theta,\phi) \, {}_sY_{l' m'}^*(\theta,\phi)=\delta_{l l'}\delta_{mm'}\,.
\eeq
We will use only the contributions due to $l=2,3,4,5,6$, for which useful information are summarized in the Table \ref{MSTsols} below.

%
%
\begin{table*}
\caption{\label{MSTsols} The generalized angular momentum parameter $\nu^l$, the (constant) Wronskian $W_{l m \omega}$ and the transmission coefficient $C_{l m \omega}^{\rm trans}$ of the MST type solutions of the homogeneous Teukolsky equation for $s=-2$ and various values of $l=2,3,4,5,6$. Here ${\mathcal L}=\ln(4 M \omega)$.}
\begin{ruledtabular}
\begin{tabular}{ll}
$\nu^{l=2}$ & $
 2 - \frac{214}{105} M^2 \omega^2  - \frac{3390466}{1157625} M^4 \omega^4  - \frac{
  153440219802466}{15021833990625} M^6\omega^6 - \frac{
  71638806585865707261481}{1520451676706008921875} M^8\omega^8 $\\
$W_{2 m \omega}$ &$\frac{\left(\frac{107 i {\mathcal L}}{56}+\frac{25 i \pi ^2}{16}-\frac{15
   \gamma  \pi }{4}+\frac{85 \pi }{16}-\frac{15 i \gamma
   ^2}{8}+\frac{809 i \gamma }{112}-\frac{333157 i}{47040}\right)
   \omega }{M^2}$\\
&$+\frac{\omega ^2 \left(-\frac{107}{28} i \pi 
   {\mathcal L}+\frac{107 \gamma  {\mathcal L}}{28}-\frac{1819 {\mathcal L}}{336}+\frac{25 \zeta
   (3)}{2}-\frac{5 i \pi ^3}{8}+\frac{25 \gamma  \pi
   ^2}{8}-\frac{1691 \pi ^2}{672}+\frac{15}{4} i \gamma ^2 \pi
   -\frac{809 i \gamma  \pi }{56}+\frac{333157 i \pi
   }{23520}-\frac{5 \gamma ^3}{4}+\frac{1023 \gamma
   ^2}{112}-\frac{460487 \gamma
   }{23520}-\frac{490159}{31360}\right)}{M}$\\
&$+\frac{15 i}{16 M^4
   \omega }+\frac{-\frac{85}{32}+\frac{15 \gamma }{8}-\frac{15 i
   \pi }{8}}{M^3}$\\
$C_{2 m \omega}^{\rm trans}$& $-2 i\omega^3 
+ M\omega^4 \left(\frac23(-1 + 3i\pi) + 
    4{\mathcal L}\right) 
+ M^2\omega^5\left(\frac{
     1340 i + 5964 \pi - 2205 i \pi^2 }{2205} - 
    \frac43 (i + 3\pi) {\mathcal L} + 
    4 i {\mathcal L}^2\right)$\\
\hline
$\nu^{l=3}$ & $3 -  \frac{26}{21} M^2\omega^2  -  \frac{21842}{33957} M^4\omega^4  -  
 \frac{381415329076}{481821815475} M^6\omega^6   -  
 \frac{47254211021655226}{35059764403038375} M^8\omega^8 $\\
$W_{3 m \omega}$ &$\frac{\frac{195 i  {\mathcal L}}{8}+\frac{525 i \pi ^2}{16}-\frac{315 \gamma 
   \pi }{4}+\frac{3507 \pi }{16}-\frac{315 i \gamma
   ^2}{8}+\frac{3897 i \gamma }{16}-\frac{799481
   i}{2240}}{M^3}$\\
&$+\frac{\omega  \left(-\frac{195}{4} i \pi 
    {\mathcal L}+\frac{195 \gamma  {\mathcal L}}{4}-\frac{2171  {\mathcal L}}{16}+\frac{525 \zeta
   (3)}{2}-\frac{105 i \pi ^3}{8}+\frac{525 \gamma  \pi
   ^2}{8}-\frac{5065 \pi ^2}{32}+\frac{315}{4} i \gamma ^2 \pi
   -\frac{3897 i \gamma  \pi }{8}+\frac{799481 i \pi
   }{1120}-\frac{105 \gamma ^3}{4}+\frac{4287 \gamma
   ^2}{16}-\frac{951451 \gamma
   }{1120}+\frac{24232057}{67200}\right)}{M^2}$\\
&$+\frac{315 i}{16
   M^5 \omega ^2}+\frac{-\frac{3507}{32}+\frac{315 \gamma
   }{8}-\frac{315 i \pi }{8}}{M^4 \omega }$\\
$C_{3 m \omega}^{\rm trans}$& $-2\omega^3 
+M\omega^4 \left(\frac23 (2 i + 3 \pi) - 4 i{\mathcal L}\right)  
+M^2\omega^5 \left(-\frac{-7 + 54 i\pi + 21 \pi^2}{21}
         + 
    \frac43i  (2 i + 3\pi) {\mathcal L} + 
    4 {\mathcal L}^2 \right)$\\
\hline
$\nu^{l=4}$ & $4 -  \frac{3142}{3465} M^2\omega^2  -  
 \frac{136964836738}{540820405125} M^4\omega^4  -  
 \frac{13932003344124287414}{84411749090784740625} M^6\omega^6  -  
 \frac{6515321108662855725628955741}{44795217188455810394959265625} M^8 \
\omega^8$\\
$W_{4 m \omega}$ &$\frac{6615 i}{16 M^6 \omega ^3}+\eta ^3 \left(\frac{6615 \gamma
   }{8 M^5 \omega
   ^2}-\frac{48069}{16 M^5
   \omega ^2}-\frac{6615 i \pi
   }{8 M^5 \omega ^2}\right)+\eta^6\left(-\frac{6615 i \gamma ^2}{8 M^4
   \omega }+\frac{280875 i
   \gamma }{44 M^4 \omega
   }-\frac{1368022661 i}{116160
   M^4 \omega }+\frac{11025 i
   \pi ^2}{16 M^4 \omega
   }-\frac{6615 \gamma  \pi }{4
   M^4 \omega }+\frac{48069 \pi
   }{8 M^4 \omega }\right.$\\
   &$\left.+\frac{32991
   i \mathcal{L}}{88 M^4 \omega
   }\right)+\eta^9\left(\frac{11025 \zeta (3)}{2
   M^3}+\left(-\frac{32991 i \pi
   }{44 M^3}+\frac{32991 \gamma
   }{44 M^3}-\frac{1198673}{440
   M^3}\right)
   \mathcal{L}\right.$\\
   &$\left.-\frac{2205 \gamma
   ^3}{4 M^3}+\frac{594741
   \gamma ^2}{88
   M^3}-\frac{1526247497 \gamma
   }{58080
   M^3}+\frac{10187674229}{43560
   0 M^3}-\frac{2205 i \pi ^3}{8
   M^3}+\frac{11025 \gamma  \pi
   ^2}{8 M^3}-\frac{815283 \pi
   ^2}{176 M^3}+\frac{6615 i
   \gamma ^2 \pi }{4
   M^3}-\frac{280875 i \gamma 
   \pi }{22
   M^3}+\frac{1368022661 i \pi
   }{58080 M^3}\right)$\\
$C_{4 m \omega}^{\rm trans}$& $2 i \eta ^3 \omega ^3+\eta ^6 \left(-4 M \omega ^4
   \mathcal{L}+\frac{2}{5} (4-5
   i \pi ) M \omega ^4\right)+\eta ^9 \left(\frac{i
   \left(-473391+3344110 i \pi
   +1334025 \pi ^2\right) M^2
   \omega ^5}{1334025}-4 i M^2
   \omega ^5
   \mathcal{L}^2+\frac{4}{5} (5
   \pi +4 i) M^2 \omega ^5
   \mathcal{L}\right)$\\
\hline
$\nu^{l=5}$ & $5-\frac{1546 M^2
   \omega ^2}{2145}-\frac{2934884558 M^4 \omega ^4}{23028130125}-\frac{5931370713515016362 M^6 \omega
   ^6}{113475665579372971875}-\frac{20361021135584219709260016691 M^8 \omega
   ^8}{708286794070019020577861765625}$\\
$W_{5 m \omega}$ & $\frac{155925 i}{16 M^7 \omega ^4}+\eta ^3 \left(\frac{155925 \gamma }{8 M^6 \omega
   ^3}-\frac{666765}{8 M^6 \omega ^3}-\frac{155925 i \pi }{8 M^6 \omega ^3}\right)+\eta
   ^6 \left(-\frac{155925 i \gamma ^2}{8 M^5 \omega ^2}+\frac{18066375 i \gamma }{104 M^5
   \omega ^2}-\frac{310732642263 i}{832832 M^5 \omega ^2}+\frac{259875 i \pi ^2}{16 M^5
   \omega ^2}-\frac{155925 \gamma  \pi }{4 M^5 \omega ^2}\right.$\\
&$\left.+\frac{666765 \pi }{4 M^5 \omega
   ^2}+\frac{730485 i \log (4 M \omega )}{104 M^5 \omega ^2}\right)+\eta ^9
   \left(-\frac{51975 \gamma ^3}{4 M^4 \omega }+\frac{4699215 \gamma ^2}{26 M^4 \omega
   }-\frac{335747175807 \gamma }{416416 M^4 \omega }+\frac{27819241507249}{29149120 M^4
   \omega }-\frac{51975 i \pi ^3}{8 M^4 \omega }+\frac{259875 \gamma  \pi ^2}{8 M^4
   \omega }\right.$\\
&$\left.-\frac{6858045 \pi ^2}{52 M^4 \omega }+\frac{155925 i \gamma ^2 \pi }{4 M^4
   \omega }-\frac{18066375 i \gamma  \pi }{52 M^4 \omega }+\frac{310732642263 i \pi
   }{416416 M^4 \omega }+\left(\frac{730485 \gamma }{52 M^4 \omega }-\frac{3123693}{52
   M^4 \omega }-\frac{730485 i \pi }{52 M^4 \omega }\right) \log (4 M \omega
   )+\frac{259875 \zeta (3)}{2 M^4 \omega }\right)$\\
$C_{5 m \omega}^{\rm trans}$& $2 \eta ^3 \omega ^3+\eta ^9 \left(\frac{\left(-54252+342160 i \pi +139425 \pi ^2\right)
   M^2 \omega ^5}{139425}-4 M^2 \omega ^5 \log ^2(4 M \omega )+\frac{4}{15} (13-15 i \pi
   ) M^2 \omega ^5 \log (4 M \omega )\right)$\\
&$+\eta ^6 \left(4 i M \omega ^4 \log (4 M
   \omega )-\frac{2}{15} (15 \pi +13 i) M \omega ^4\right)$\\
\hline
$\nu^{l=6}$ & $6-\frac{1802 M^2 \omega ^2}{3003}-\frac{169547733896 M^4 \omega
   ^4}{2301891887295}-\frac{82446690884152932671 M^6 \omega
   ^6}{3944111533764126174450}-\frac{18568771026973098936173491476839 M^8 \omega
   ^8}{2343648850762937006487915050934600}$\\
$W_{6 m \omega}$ &$\frac{66891825 i}{256 M^8 \omega ^5}+\eta ^3 \left(\frac{66891825 \gamma }{128 M^7 \omega
   ^4}-\frac{2568009015}{1024 M^7 \omega ^4}-\frac{66891825 i \pi }{128 M^7 \omega
   ^4}\right)+\eta ^6 \left(-\frac{66891825 i \gamma ^2}{128 M^6 \omega
   ^3}+\frac{2648288115 i \gamma }{512 M^6 \omega ^3}-\frac{4626372940281 i}{372736 M^6
   \omega ^3}+\frac{111486375 i \pi ^2}{256 M^6 \omega ^3}\right.$\\
&$\left. -\frac{66891825 \gamma  \pi
   }{64 M^6 \omega ^3}+\frac{2568009015 \pi }{512 M^6 \omega ^3}+\frac{20069775 i \log (4
   M \omega )}{128 M^6 \omega ^3}\right)+\eta ^9 \left(-\frac{22297275 \gamma ^3}{64 M^5
   \omega ^2}+\frac{2728567215 \gamma ^2}{512 M^5 \omega ^2}-\frac{700975807443 \gamma
   }{26624 M^5 \omega ^2}+\frac{1926087053600823}{52183040 M^5 \omega ^2}\right.$\\
&$-\frac{22297275
   i \pi ^3}{128 M^5 \omega ^2}+\frac{111486375 \gamma  \pi ^2}{128 M^5 \omega
   ^2}-\frac{4119456825 \pi ^2}{1024 M^5 \omega ^2}+\frac{66891825 i \gamma ^2 \pi }{64
   M^5 \omega ^2}-\frac{2648288115 i \gamma  \pi }{256 M^5 \omega ^2}+\frac{4626372940281
   i \pi }{186368 M^5 \omega ^2}$\\
&$\left.+\left(\frac{20069775 \gamma }{64 M^5 \omega
   ^2}-\frac{5393417535}{3584 M^5 \omega ^2}-\frac{20069775 i \pi }{64 M^5 \omega
   ^2}\right) \log (4 M \omega )+\frac{111486375 \zeta (3)}{32 M^5 \omega ^2}\right)$\\
$C_{6 m \omega}^{\rm trans}$& $-2 i \eta ^3 \omega ^3+\eta ^9 \left(\frac{\left(288028 i+1671516 \pi -693693 i \pi
   ^2\right) M^2 \omega ^5}{693693}+4 i M^2 \omega ^5 \log ^2(4 M \omega )-\frac{4}{21}
   (21 \pi +19 i) M^2 \omega ^5 \log (4 M \omega )\right)$\\
&$+\eta ^6 \left(4 M \omega ^4
   \log (4 M \omega )+\frac{2}{21} (-19+21 i \pi ) M \omega ^4\right)$\\
\end{tabular}
\end{ruledtabular}
\end{table*}

Integrating Eq. \eqref{eq_dE_rad} over $t$ or over $\omega$ one finds the relations
\bea
E_{\rm rad}&=&\int dt \frac{dE_{\rm rad}}{dt}=\int d\omega \frac{dE}{d\omega}\,,
\eea
that is
\bea
E_{\rm rad} 
&=& \frac{1}{4\pi}\sum_{l ,m}\int \frac{d\omega}{2\pi}  \frac{|\widehat Z_{l m \omega}|^2}{ \omega^2}\,, 
\eea
and then
\beq
\frac{dE}{d\omega}=\frac{1}{4\pi}\sum_{l ,m} \frac{1}{2\pi}  \frac{|\widehat Z_{l m \omega}|^2}{ \omega^2}=\sum_l \frac{dE}{d\omega}\Big|^{(l)}\,.
\eeq
Explicitly, for example  
\bea
\frac{dE}{d\omega}\Big|^{(2)}=\frac{\sum_{m=- 2}^2 |\widehat Z_{2 m \omega}|^2 
}{8\pi^2 \omega^2}\,,
\eea
etc., implying
    \bea
\frac{M^2}{\mu^2}\frac{dE}{d t}\Big|_{l=2}&=&\frac{512 \alpha ^{10}}{885735
   t^{10/3}}+\frac{45056 \alpha
   ^{12}}{18600435 t^4}\nn\\
   &+&\frac{\alpha
   ^{13}}{t^{13/3
   }}\left(\frac{1024 \left(41+10
   \sqrt{3} \pi \right) }{71744535}\right.\nn\\
   &+&\left.\frac{4096
   \mathcal{L}}{4782969}\right)+\frac{991232 \alpha
   ^{14}}{390609135 t^{14/3}}\nn\\
   &+&\frac{\alpha
   ^{15}}{t^5}\left(\frac{720896 
   \mathcal{L}}{167403915}-\frac{
   45056 \left(4 \sqrt{3} \pi
   -273\right)}{1506635235}\right)\qquad\nn\\
   &+&\frac{\alpha^{16}}{t^{16/3}}\Big[-\frac{556652013056}{4698441980347
   5}\nn\\
   &-&\frac{3303424 \pi }{13559717115
   \sqrt{3}}+\frac{137216 \pi ^2}{129140163}\nn\\
   &+&\left(\frac{180224}{71744535}+\frac{6553
   6 \pi }{129140163 \sqrt{3}}\right)\gamma\nn\\
   &-&\frac{65536 \gamma ^2}{129140163}+\left(\frac{170528768}{40679151345}\right.\nn\\
   &-&\left.\frac{131072 \gamma
   }{129140163}+\frac{57344 \pi
   }{43046721 \sqrt{3}}\right)\mathcal{L}\nn\\
   &+&\frac{40960
   \mathcal{L}^2}{129140163}\Big]+O\left(t^{-\frac{17}{3}}\right)\,,
\eea
\bea
\frac{M^2}{\mu^2}\frac{dE}{d t}\Big|_{l=3}&=&\frac{18496000 \alpha
   ^{16}}{73222472421 t^{16/3}}+O\left(t^{-\frac{17}{3}}\right)\,,\qquad
\eea
\bea
\frac{M^2}{\mu^2}\frac{dE}{d t}\Big|_{l=4}&=&\frac{40960 \alpha
   ^{14}}{18983603961 t^{14/3}}\nn\\
   &+&\frac{\frac{327680 \alpha ^{17}
   \mathcal{L}}{73222472421}-\frac
   {16384 \left(70 \sqrt{3} \pi
   -719\right) \alpha
   ^{17}}{1537671920841}}{t^{17/3}
   }\nn\\
   &-&\frac{518193152 \alpha
   ^{18}}{13155637544973 t^6}+O\left(t^{-\frac{19}{3}}\right)\,,
\eea
\bea
\frac{M^2}{\mu^2}\frac{dE}{d t}\Big|_{l=5}&=&\frac{17920 \alpha
   ^{16}}{115063885233 t^{16/3}}\nn\\
   &+&\frac{5820416 \alpha
   ^{18}}{4487491524087 t^6}\nn\\
   &+&\frac{\frac{4096 \left(140
   \sqrt{3} \pi -197\right) \alpha
   ^{19}}{9320174703873}+\frac{114
   6880 \alpha ^{19}
   \mathcal{L}}{3106724901291}}{t^
   {19/3}}\nn\\
   &+&\frac{2363088896 \alpha
   ^{20}}{875060847196965
   t^{20/3}}\nn\\
   &+&\frac{\frac{46563328 \alpha ^{21}
   \mathcal{L}}{13462474572261}-\frac{1662976 \left(70 \sqrt{3}
   \pi -3321\right) \alpha
   ^{21}}{1817434067255235}}{t^7}\nn\\
   &+&O\left(t^{-\frac{22}{3}}\right)\,,
\eea
\bea
\frac{M^2}{\mu^2}\frac{dE}{d t}\Big|_{l=6}&=&\frac{28772925440 \alpha
   ^{22}}{107955583594960959
   t^{22/3}}\nn\\
   &+&O\left(t^{-\frac{23}{3}}\right)\,,
\eea
where 
\beq
{\mathcal L}=\ln \left(\frac{4M}{3\sqrt{3}t}\right)\,.
\eeq
Summing up all these contributions one gets the final result
\begin{widetext}
\bea
\label{sum_l_t}
\frac{M^2}{\mu^2}\frac{dE}{dt}&=&\frac{M^2}{\mu^2}\sum_{l=2}^{6} \frac{dE}{d t}\Big|_{l}\nonumber\\
&=& \frac{32\cdot 2^{2/3}}{405 \cdot3^{1/3} T^{10/3}}
+\frac{2816 }{2835 T^4}+\frac{2^{2/3}}{3^{1/3}T^{13/3}}\left(\frac{1312}{3645}+\frac{64 \pi }{243\sqrt{3}}+\frac{128\mathcal{L}}{243}\right)+\frac{1076224
   \cdot2^{1/3}}{229635\cdot 3^{2/3}T^{14/3}}\qquad\nn\\
   &+&\left(\frac{18304}{1215}-\frac{5632 \pi }{8505 \sqrt{3}}+\frac{22528 \mathcal{L}}{2835}\right)\frac{1}{T^5}+\Big[-\frac{8512131704}{265228425}-\frac{51616 \pi}{76545\sqrt{3}}+\frac{2144 \pi ^2}{729}\nn\\
   &+&\left(\frac{2816}{405}+\frac{1024 \pi }{729\sqrt{3}}\right)\gamma-\frac{1024 \gamma
   ^2}{729}+\left(\frac{2664512}{229635}-\frac{2048 \gamma }{729}+\frac{896 \pi }{243\sqrt{3}}\right)\mathcal{L}+\frac{640\mathcal{L}^2}{729}\Big]\frac{2^{2/3}}{3^{1/3}T^{16/3}}\nn\\
   &+&O\left(t^{-17/3}\right)
\,.
\eea
Here the accuracy of the final result is $O\left(t^{-17/3}\right)$, while  certain intermediate results have  been computed with higher accuracy. For example, the terms $O(t^{-6})$ in the contribution $l=4$, the terms $O(t^{-6})$, $O(t^{-19/3})$, $O(t^{-19/3})$, $O(t^{-20/3})$, $O(t^{-7})$ in  the contribution $l=5$ and the terms   $O(t^{-22/3})$  in  the contribution $l=6$  cannot be taken because they would require higher order computations for the term $l=2$.

\end{widetext}

In the frequency domain 
\bea
\frac{1}{\mu^2}\frac{dE}{d\omega}\Big|_{l=2}&=&\frac{12\cdot 6^{2/3}  \Gamma
   \left(\frac{4}{3}\right)^2}{5\pi }(M\omega)^{4/3}\nonumber\\
&+&\frac{176}{35} \sqrt{3} \eta ^2
   (M\omega)^2\nn\\
   &-&\frac{24}{5} 6^{2/3} 
    \Gamma
   \left(\frac{4}{3}\right)^2\eta ^3(M\omega)^{7/3}\nonumber\\
&-&\frac{352}{35} \sqrt{3} \pi 
   \eta ^5 (M\omega)^3\qquad\nn\\
   &+&\frac{8712 \cdot 6^{1/3}  \Gamma
   \left(\frac{2}{3}\right)^2}{24
   5 \pi }\eta ^4
   (M\omega)^{8/3}\nn\\
&+&\frac{\Gamma(4/3)^2}{80 \cdot6^{1/3} \pi }\Big[7488 \log ^2(4 M \omega )\nn\\
&+&(14976
   \gamma -23376) \log (4 M \omega
   )+432 \pi ^2+7488 \gamma
   ^2\nn
   \eea
   \bea
   &-&23376 \gamma +18493-2880
   \psi
   ^{(1)}\left(\frac{11}{3}\right)\Big]\eta^6(M \omega)^{10/3}\nn\\
   &+&O(\omega^{11/3})\,,
\eea

\beq
\frac{1}{\mu^2}\frac{dE}{d\omega}\Big|_{l=3}=\frac{1445\cdot 2^{2/3}  \Gamma
   \left(\frac{4}{3}\right)^2}{63
   \cdot3^{1/3} \pi }\eta ^4
   (M\omega)^{10/3}+O(\omega^{11/3})\,,
\eeq
\bea
\frac{1}{\mu^2}\frac{dE}{d\omega}\Big|_{l=4}&=&\frac{40 \cdot2^{1/3} 
   \Gamma
   \left(\frac{2}{3}\right)^2}{441\cdot 3^{2/3} \pi }(M\omega)^{8/3}\nonumber\\
&-& \frac{80 \cdot2^{1/3}  \Gamma
   \left(\frac{2}{3}\right)^2}{441\cdot 3^{2/3}}\eta ^3
   (M\omega)^{11/3}\nn\\
   &+&\frac{63256}{56595
   \sqrt{3}}\eta ^4 (M\omega)^4+O(\omega^{13/3})\,,\qquad
\eea
\bea
\frac{1}{\mu^2}\frac{dE}{d\omega}\Big|_{l=5}&=&\frac{7\cdot 2^{2/3}  \Gamma
   \left(\frac{4}{3}\right)^2}{49
   5 \cdot3^{1/3} \pi }(M\omega)^{10/3}\nonumber\\
&+&\frac{812}{19305
   \sqrt{3}}\eta ^2 (M\omega)^4\nn\\
   &-&\frac{14\cdot 2^{2/3}  \Gamma
   \left(\frac{4}{3}\right)^2}{49
   5 \cdot3^{1/3}}\eta ^3
   (M\omega)^{13/3}\nn\\
   &+&\frac{11774 \cdot2^{1/3}  \Gamma
   \left(\frac{2}{3}\right)^2}{27
   885\cdot 3^{2/3} \pi }\eta ^4
   (M\omega)^{14/3}\nn\\
   &-&\frac{1624 \pi  }{19305 \sqrt{3}}\eta ^5
   (M\omega)^5+O(\omega^{16/3})\,,\nn\\
\eea
\bea
\frac{1}{\mu^2}\frac{dE}{d\omega}\Big|_{l=6}&=&\frac{43904\cdot 2^{2/3}  \Gamma
   \left(\frac{4}{3}\right)^2}{63
   7065 \cdot3^{1/3} \pi }\eta ^4
   (M\omega)^{16/3}\nn\\
   &+&O(\omega^{17/3})\,.\qquad
\eea
Setting $\eta=1$, the final results is given by
\begin{widetext}

\bea
\label{sum_l_omega}
\frac{1}{\mu^2}\frac{dE}{d\omega}&=&\frac{1}{\mu^2}\sum_{l=2}^6\frac{dE}{d\omega}\Big|_{l}\nonumber\\
&=& \frac{12\cdot 6^{2/3} \Gamma
   \left(\frac{4}{3}\right)^2}{5
   \pi }(M\omega)^\frac{4}{3}+\frac{176 \sqrt{3}}{35}(M\omega)^2-\frac{24}{5} 6^{2/3} \Gamma
   \left(\frac{4}{3}\right)^2(M\omega)^\frac{7}{3}+\frac{33632 \cdot2^{1/3} \Gamma
   \left(\frac{2}{3}\right)^2}{31
   5\cdot 3^{2/3} \pi }(M\omega)^\frac{8}{3}\nonumber\\
&-&\frac{352 \sqrt{3} \pi }{35}(M\omega)^3+\frac{\Gamma
   \left(\frac{4}{3}\right)^2 }{6160
   \cdot6^{1/3} \pi }\Big[576576 \log ^2(4 M \omega
   )+(1153152 \gamma -1799952)
   \log (4 M \omega )\nn\\
   &+&33264 \pi
   ^2+576576 \gamma ^2-1799952
   \gamma +1706713-221760 \psi
   ^{(1)}\left(\frac{11}{3}\right)\Big](M
   \omega )^{10/3}+O\left(\omega^{11/3}\right)\,,
\eea
\end{widetext}
and again the final accuracy is $O\left(\omega^{11/3}\right)$ in spite of the fact that certain contributions have been computed with higher accuracy, as in the time domain counterpart of this calculation.

\section{Concluding Remarks}

We have computed analytically the energy emitted by a (either  scalar or massive)  particle  radially falling   in the Schwarzschild black hole spacetime, within the framework of gravitational self force, at the first self force accuracy level. The results, presented in Post-Newtonian expanded form, are shown here for the first time:
See Eq. \eqref{E_rad_scal} for the scalar case and Eqs. \eqref{sum_l_t} and \eqref{sum_l_omega} for the gravitational case.
Their range of validity is limited by the Post-Newtonian assumptions, i.e., break down as soon as the perturbing particle enters the strong field region.
All the computational details are made explicit here. These results 1) have an intrinsic interest, 2) can be informative of other formalisms like the Effective-One-Body (EOB) formalism and 3) can be used as bed test for numerical calculations. 

This work is also preliminar  to forthcoming analytical study of the same problem but in the strong-field regime. The leading contribution to the power emitted at infinity in the limit $GM\omega \gg 1$ was computed in \cite{Zerilli:1970wzz} and exhibits a decreasing exponential behavior, $e^{-4\pi M\omega}$. It would be  interesting to develop a systematic procedure to compute various subleading orders in the large-$GM\omega$ expansion. The ultimate goal would be to match the weak- and strong-field results in their overlapping region. Indeed, the radial fall is peculiar in the sense that at a certain time $t_{\rm max}$ from the beginning of the process one losses the possibility to study the problem within a post-Newtonian framework. In other words post-Newtonian analysis stops at $t_{\rm max}$ (i.e., a natural  length-scale, with a corresponding $\omega_{\rm max}$ in the Fourier domain, associated with the approach) when the particle enters the strong field region.
Similarly, a study of the orbit close to the horizon can be valid up to certain distance from the horizon itself, marked by another length-scale. Matching of the two scales within a more general approach (when available) will give a consistent solution of the problem. As shown in \cite{Davis:1971gg}, and as can be readily anticipated from the leading behaviors $\omega^{4/3}$ at $\omega\sim 0$ and $e^{-4\pi M\omega}$ at $\omega\gg 1$, the power spectrum exhibits a maximum in this overlap region. The aim of this future project is to provide a new analytical perspective for addressing this class of physical problems, which have traditionally been treated primarily through numerical methods \cite{Bini:2026ova}. 

On the other hand, this type of computation could be extended to other classes of gravitational solutions, generally in dimensions higher than four, which admit the presence of regular caps in place of an event horizon. Such geometries appear as black hole mimickers, whose underlying philosophy is to address the well-known black hole information paradox \cite{Lunin:2001jy}. Within this broad framework, of particular interest are certain five-dimensional Einstein-Maxwell solutions, dubbed Topological Stars \cite{Bah:2020pdz,Heidmann:2025pbb,Bianchi:2025uis}, as well as the W-soliton'' \cite{Dima:2025tjz}. These backgrounds have already been employed to study scalar perturbations \cite{Bianchi:2024vmi,Bianchi:2024rod,Fucito:2024wlg,DiRusso:2025lip,Bianchi:2025aei,Bini:2025ltr,Bianchi:2025ydq,Bini:2025bll,Cipriani:2025ikx}, and can be further developed by completing the analysis with the study of radial infall, which indeed we plan to address in future work.

\appendix
\begin{widetext}
\section{MST solutions}\label{MSTapp}
In this appendix, we display the PN expansion (up to $O(\eta^{10})$ of the MST \lq\lq in" and \lq\lq up" solutions for $l=2$, derived from Eqs. (101), (102) and (134) of Ref. \cite{Sasaki:2003xr}.

\bea
    R_{2 \omega}^{\rm in}(r)&=&\frac{r^4}{16 M^4}+\frac{i r^5 \omega }{24 M^4}\eta+\left(-\frac{11 r^6 \omega ^2}{672
   M^4}-\frac{r^3}{4 M^3}\right)\eta^2+\left(-\frac{i r^7 \omega ^3}{224
   M^4}-\frac{7 i r^4 \omega }{96
   M^3}\right)\eta^3+\left(\frac{23 r^8 \omega ^4}{24192
   M^4}-\frac{17 r^5 \omega
   ^2}{504 M^3}+\frac{r^2}{4 M^2}\right)\eta^4\qquad\nn\\
   &+&\left(\frac{i r^9 \omega ^5}{6048
   M^4}-\frac{113 i r^6 \omega
   ^3}{4032 M^3}-\frac{i r^3
   \omega }{12 M^2}\right)\eta^5+\left(-\frac{13 r^{10} \omega
   ^6}{532224 M^4}+\frac{127 r^7
   \omega ^4}{12096
   M^3}+\frac{107 r^4 \omega ^2
   \log \left(\frac{2
   M}{r}\right)}{840
   M^2}+\frac{24253 r^4 \omega
   ^2}{47040 M^2}\right.\nn\\
   &-&\left.\frac{\pi ^2
   r^4 \omega ^2}{24 M^2}\right)\eta^6+\left(-\frac{5 i r^{11} \omega
   ^7}{1596672 M^4}+\frac{43 i
   r^8 \omega ^5}{16128
   M^3}+\frac{107 i r^5 \omega ^3
   \log \left(\frac{2
   M}{r}\right)}{1260
   M^2}+\frac{1808 i r^5 \omega
   ^3}{6615 M^2}-\frac{i \pi ^2
   r^5 \omega ^3}{36 M^2}+\frac{i
   r^2 \omega }{12 M}\right)\eta^7\qquad\nn\\
   &+&\left(\frac{59 r^{12} \omega
   ^8}{166053888 M^4}-\frac{389
   r^9 \omega ^6}{739200
   M^3}-\frac{23971 r^6 \omega
   ^4}{282240 M^2}+\frac{11 \pi
   ^2 r^6 \omega ^4}{1008
   M^2}+\left(-\frac{1177 r^6
   \omega ^4}{35280
   M^2}-\frac{107 r^3 \omega
   ^2}{210 M}\right) \log
   \left(\frac{2 
   M}{r}\right)\right.\nn\\
   &-&\left.\frac{1355 r^3
   \omega ^2}{1176 M}+\frac{\pi
   ^2 r^3 \omega ^2}{6 M}\right)\eta^8+\left(\frac{i r^{13} \omega
   ^9}{27675648 M^4}-\frac{6803 i
   r^{10} \omega ^7}{79833600
   M^3}-\frac{187 i r^7 \omega
   ^5}{10368 M^2}+\frac{i \pi ^2
   r^7 \omega ^5}{336
   M^2}\right.\nn\\
   &+&\left.\left(-\frac{107 i r^7
   \omega ^5}{11760
   M^2}-\frac{107 i r^4 \omega
   ^3}{720 M}\right) \log
   \left(\frac{2
   M}{r}\right)+\frac{185053 i
   r^4 \omega ^3}{120960
   M}-\frac{61 i \pi ^2 r^4
   \omega ^3}{1680 M}-\frac{i r^4
   \omega ^3 \zeta (3)}{M}\right)\eta^9+O(\eta^{10})\,,\nn\\
\eea
\bea
R_{2 \omega}^{\rm up}(r)&=&-\frac{3 i}{r \omega }-3\eta+\left(\frac{3 i r \omega }{2}-\frac{3 i
   M}{r^2 \omega }\right)\eta^2+\left(-\frac{6 \gamma  M}{r}+\frac{2
   M}{r}+\frac{6 i \pi 
   M}{r}+\frac{r^2 \omega ^2}{2}\right)\eta^3+\left(-\frac{24 i M^2}{7 r^3 \omega }+6
   i \gamma  M \omega \right.\nn\\
   &-&\left.\frac{13 i
   M \omega }{2}+6 \pi  M \omega
   -\frac{1}{8} i r^3 \omega ^3\right)\eta^4+\left(\gamma  \left(3 M r \omega
   ^2-\frac{6
   M^2}{r^2}\right)-\frac{12
   M^2}{7 r^2}+\frac{6 i \pi 
   M^2}{r^2}-6 M r \omega ^2-3 i
   \pi  M r \omega ^2\right.\nn\\
   &+&\left.\frac{31
   r^4 \omega ^4}{40}\right)\eta^5+\left(-\frac{30 i M^3}{7 r^4 \omega
   }+\gamma  \left(-\frac{354 i
   M^2 \omega }{35 r}+\frac{12
   \pi  M^2 \omega }{r}-i M r^2
   \omega ^3\right)-\frac{214 i
   M^2 \omega  \log (2  r
   \omega )}{35 r}\right.\nn\\
   &+&\left.\frac{6 i
   \gamma ^2 M^2 \omega
   }{r}+\frac{1265 i M^2 \omega
   }{49 r}-\frac{7 i \pi ^2 M^2
   \omega }{r}-\frac{4 \pi  M^2
   \omega }{r}+\frac{31}{8} i M
   r^2 \omega ^3-\pi  M r^2
   \omega ^3+\frac{43}{80} i r^5
   \omega ^5\right)\eta^6\nn\\
   &+&\left(-\frac{48 M^3}{7 r^3}+\frac{48 i
   \pi  M^3}{7
   r^3}-\frac{214}{35} M^2 \omega
   ^2 \log (2  r \omega )+6
   \gamma ^2 M^2 \omega
   ^2+\frac{6976 M^2 \omega
   ^2}{245}-7 \pi ^2 M^2 \omega
   ^2+13 i \pi  M^2 \omega
   ^2\right.\nn\\
   &+&\left.\gamma  \left(-\frac{48
   M^3}{7 r^3}-\frac{669 M^2
   \omega ^2}{35}-12 i \pi  M^2
   \omega ^2-\frac{1}{4} M r^3
   \omega ^4\right)-\frac{7}{20}
   M r^3 \omega ^4+\frac{1}{4} i
   \pi  M r^3 \omega
   ^4-\frac{117}{560} r^6 \omega
   ^6\right)\eta^7\nn\\
   &+&\left(-\frac{40 i M^4}{7 r^5 \omega
   }+\frac{26899 i M^3 \omega
   }{735 r^2}-\frac{7 i \pi ^2
   M^3 \omega }{r^2}+\frac{24 \pi
    M^3 \omega }{7
   r^2}-\frac{28943 i M^2 r
   \omega ^3}{1470}+\frac{7}{2} i
   \pi ^2 M^2 r \omega ^3+12 \pi 
   M^2 r \omega ^3\right.\nn\\
   &+&\left.\gamma ^2
   \left(\frac{6 i M^3 \omega
   }{r^2}-3 i M^2 r \omega
   ^3\right)+\left(-\frac{214 i
   M^3 \omega }{35
   r^2}+\frac{107}{35} i M^2 r
   \omega ^3+\frac{8}{5} i M r^4
   \omega ^5\right) \log (2  
   r \omega )\right.\nn\\
   &+&\left.\gamma 
   \left(-\frac{94 i M^3 \omega
   }{35 r^2}+\frac{12 \pi  M^3
   \omega }{r^2}+\frac{527}{35} i
   M^2 r \omega ^3-6 \pi  M^2 r
   \omega ^3+\frac{1}{20} i M r^4
   \omega
   ^5\right)-\frac{313}{400} i M
   r^4 \omega ^5-\frac{3}{4} \pi 
   M r^4 \omega ^5\right.\nn\\
   &-&\left.\frac{769 i
   r^7 \omega ^7}{13440}\right)\eta^8+\left(-\frac{325 M^4}{21 r^4}+\frac{60
   i \pi  M^4}{7 r^4}+\frac{4
   \gamma ^3 M^3 \omega
   ^2}{r}+\frac{907 M^3 \omega
   ^2}{2205 r}+\frac{6 i \pi ^3
   M^3 \omega ^2}{r}+\frac{92 \pi
   ^2 M^3 \omega ^2}{35
   r}\right.\nn\\
   &-&\left.\frac{2530 i \pi  M^3
   \omega ^2}{49 r}+\frac{8 M^3
   \omega ^2 \zeta
   (3)}{r}-\frac{52673 M^2 r^2
   \omega ^4}{4410}+\frac{7}{6}
   \pi ^2 M^2 r^2 \omega
   ^4-\frac{31}{4} i \pi  M^2 r^2
   \omega ^4+\gamma ^2
   \left(-\frac{568 M^3 \omega
   ^2}{35 r}\right.\right.\nn\\
   &-&\left.\left.\frac{12 i \pi  M^3
   \omega ^2}{r}-M^2 r^2 \omega
   ^4\right)+\left(-\frac{428
   \gamma  M^3 \omega ^2}{35
   r}+\frac{428 M^3 \omega
   ^2}{105 r}+\frac{428 i \pi 
   M^3 \omega ^2}{35
   r}+\frac{107}{105} M^2 r^2
   \omega ^4\right.\right.\nn\\
   &-&\left.\left.\frac{16}{15} M r^5
   \omega ^6\right) \log (2 
   r \omega )+\gamma 
   \left(-\frac{60 M^4}{7
   r^4}+\frac{40946 M^3 \omega
   ^2}{735 r}-\frac{14 \pi ^2 M^3
   \omega ^2}{r}+\frac{708 i \pi 
   M^3 \omega ^2}{35
   r}+\frac{3683}{420} M^2 r^2
   \omega ^4\right.\right.\nn\\
   &+&\left.\left.2 i \pi  M^2 r^2
   \omega ^4+\frac{1}{120} M r^5
   \omega ^6\right)+\frac{797 M
   r^5 \omega
   ^6}{3150}-\frac{13}{24} i \pi 
   M r^5 \omega ^6+\frac{1471 r^8
   \omega ^8}{120960}\right)\eta^9+O\left(\eta^{10}\right)\,.
\eea
\end{widetext}

\section*{Acknowledgments}

We thank M.~Bianchi, D.~Fioravanti, F.~Fucito, A.~Geralico, T.~Manton, J.~F.~Morales, M.~Rossi, A.~Tokareva  for useful discussions. D.~B.  acknowledges membership to the Italian Gruppo Nazionale per la Fisica Matematica (GNFM) of the Istituto Nazionale di Alta Matematica
(INDAM). 
  
\section{Data availability} 
The data that support the findings of this article are openly available \cite{dataval}.

\end{document}